\documentstyle[11pt,aaspp4,graphicx,epsfig,flushrt]{article}
\begin{document}
\title{Roles of Hyperons in Neutron Stars}
%\vskip 2.0cm
\author{Shmuel Balberg, Itamar Lichtenstadt}
\vskip 0.5cm
\affil{\small The Racah Institute of Physics, 
The Hebrew University, Jerusalem 91904, Israel}
\vskip 0.75cm
\and
\vskip 0.75cm
\author{Gregory. B. Cook}
\vskip 0.5cm
\affil{\small Center for Radiophysics and Space Research, 
Space Sciences Building,\\ 
\vskip 0.25cm
Cornell University, Ithaca, NY 14853}
\vskip 1.5cm

\date{today}
%\begin{document}
\setcounter{page}{1}

%=================================================================
%    ABSTRACT
%=================================================================
\begin{abstract}

We examine the roles the presence of hyperons in the cores of neutron stars 
may play in determining global properties of these stars. The study is based 
on estimates that hyperons appear in neutron star matter at about twice the 
nuclear saturation density, and emphasis is placed on effects that can be 
attributed to the general multi-species composition of the matter, 
hence being only weakly dependent on the specific modeling of strong 
interactions. Our analysis indicates that hyperon formation not only 
softens the equation of state but also severely constrains its values at 
high densities. Correspondingly, the valid range for the 
maximum neutron star mass is limited to about $1.5\!-\!1.8\;M_\odot$, 
which is a much narrower range than available when hyperon formation is 
ignored. Effects concerning neutron star radii and rotational evolution 
are suggested, and we demonstrate that the effect of hyperons on the 
equation of state allows a reconciliation of observed pulsar glitches with a 
low neutron star maximum mass. We discuss the effects hyperons may 
have on neutron star cooling rates, including recent results which indicate 
that hyperons may also couple to a superfluid state in high density matter. 
We compare nuclear matter to matter with hyperons and show that once 
hyperons accumulate in neutron star matter they reduce the likelihood of a 
meson condensate, but increase the susceptibility to baryon deconfinement, 
which could result in a mixed baryon-quark matter phase. 
  
\end{abstract}

\newpage
\setlength{\topmargin}{-1.3in}
\renewcommand{\baselinestretch}{1.1}
%=================================================================
%    INTRODUCTION
%=================================================================

\section{Introduction}

The existence of stable matter at supernuclear densities is unique to 
neutron stars. Unlike all other physical systems in nature, where the 
baryonic component appears in the form of atomic nuclei, matter in the 
cores of neutron stars is expected to be a homogeneous mixture of 
hadrons and leptons. As a result the macroscopic features of neutron 
stars, including some observable quantities, have the potential to
illuminate the physics of supernuclear densities. In this sense, 
neutron stars serve as cosmological laboratories for hadronic physics. 
A specific feature of supernuclear densities is the possibility for
new hadronic degrees of freedom to appear, in addition to neutrons and 
protons. One such possible degree of freedom is the formation of hyperons - 
strange baryons - which is the main subject of the present work. 
Other possible degrees of freedom include meson condensation and a 
deconfined quark phase.

While hyperons are unstable under terrestrial conditions and decay into 
nucleons through the weak interaction, the equilibrium conditions in neutron 
stars can make the reverse process, i.e., the conversion of nucleons into 
hyperons, energetically favorable. The appearance of hyperons in neutron stars 
was first suggested by Ambartsumyan \& Saakyan (1960) and has since been 
examined in many works. Earlier calculations include the works of 
Pandharipande (1971b), Bethe \& Johnson (1974) and Moszkowski (1974), which 
were performed by describing the nuclear force in Schr\"odinger theory.
In recent years, studies of high density matter with hyperons have been 
performed mainly in the framework of field theoretical models (\cite{GH}, 
\cite{WWH}, Knorren, Prakash \& Ellis 1995, \cite{SMH}, \cite{HetalH}). 
For a review, see Glendenning (1996) and Prakash et al.\ (1997). It was also 
recently demonstrated that good agreement with these models can be attained 
with an effective potential model (\cite{BGH}). 

These recent works share a wide consensus that hyperons should appear in 
neutron star (cold, beta-equilibrated, neutrino-free) matter at a density 
of about twice the nuclear saturation density. This consensus is attributed 
to the fact that all these more modern works base their estimates of 
hyperon-nucleon and hyperon-hyperon interactions on the experimental 
constraints inferred from hypernuclei. The fundamental qualitative result 
from hypernuclei experiments is that hyperon related interactions are similar 
in character and in order of magnitude to nucleon-nucleon interactions. In a 
broader sense, this result indicates that in high density matter, 
the differences between hyperons 
and nucleons will be less significant than for free particles.

The aim of the present work is to examine what roles the presence of hyperons 
in the cores of neutron stars may play in determining the global properties of 
these stars. 
%Such effects are of obvious interest, since they may be 
%constrained by observation, which thus gives insight into the physics of 
%high density matter. 
We place special emphasis on 
effects which can be attributed to the multi-species composition of the matter
while being only weakly dependent on the details
of the model used to describe the underlying strong interactions.

We begin our survey in \S~2 with a brief summary of the equilibrium conditions
which determine the formation and abundance of hyperon species in neutron star 
cores. A review of the widely accepted results 
regarding hyperon formation in neutron stars is given in \S~3. 
We devote \S~4 to an examination of the effect of hyperon 
formation on the equation of state of dense matter, and the corresponding 
effects on the star's global properties: maximum mass, mass-radius 
correlations, rotation limits, and crustal sizes. In \S~5 we discuss 
neutron star cooling rates, where hyperons might play a decisive role. 
A discussion of the effects of hyperons on phase transitions
which may occur in high density matter is given in \S~6. 
Conclusions and discussion are offered in \S~7.

%=================================================================
%    Equilibrium Conditions for Hyperon Formation Neutron Stars
%=================================================================

\section{Equilibrium Conditions for Hyperon Formation Neutron Stars}

In the following discussion we assume that the cores of neutron stars are 
composed of a mixture of baryons and leptons in full beta equilibrium (thus 
ignoring possible meson condensation and a deconfined quark phase - 
these issues will be picked up again in \S~6). The procedure for solving the 
equilibrium composition of such matter has been describes in many works 
(see e.g., Glendenning (1996) and Prakash et al.\ (1997) and references 
therein), and in essence requires chemical equilibrium of all weak processes 
of the type 
\begin{equation}\label{eq:genpro}
   B_1\rightarrow B_2+\ell+\bar{\nu}_\ell\;\;;\;\;
   B_2+\ell\rightarrow B_1+\nu_\ell\;\;,
\end{equation}
where $B_1$ and $B_2$ are baryons, $\ell$ is a lepton (electron or muon), 
and $\nu$ ($\bar{\nu}$) is its corresponding neutrino (anti-neutrino). 
Charge conservation is implied in all processes, determining the legitimate 
combinations of baryons which may couple together in such reactions.

Imposing all the conditions for chemical equilibrium yields the ground state 
composition of beta-equilibrated high density matter. The equilibrium 
composition of such matter at any given baryon density, $\rho_B$, 
is described by the relative fraction of each species of baryons 
$x_{B_i}\!\equiv\!\rho_{B_i}/\rho_B$ and leptons 
$x_\ell\!\equiv\!\rho_\ell/\rho_B$. 

Evolved neutron stars can be assumed to be 
transparent to neutrinos on any relevant time scale so that neutrinos are
absent and $\mu_\nu\!=\mu_{\bar{\nu}}\!=\!0$. All equilibrium conditions may 
then be summarized by a single generic equation 
\begin{equation}\label{eq:genequi}
    \mu_i=\mu_n-q_i\mu_e \;\;,
\end{equation}
where $\mu_i$ and $q_i$ are, respectively, the chemical potential and 
electric charge of baryon species $i$, $\mu_n$ is the neutron chemical 
potential, and $\mu_e$ is the electron chemical potential. Note that
in the absence of neutrinos, equilibrium requires $\mu_e\!=\!\mu_\mu$. 
The neutron and electron chemical potentials are 
constrained by the requirements of a constant total baryon number and 
electric charge neutrality,
\begin{equation}\label{eq:chrge0}
    \sum_i{x_{B_i}}=1\;\; ; \;\; 
    \sum_i{q_i x_{B_i}}+\sum_\ell{q_\ell x_\ell=0}\;\;.
\end{equation} 

The temperature range of evolved neutron stars is typically much lower 
than the relevant chemical potentials of baryons and leptons at 
supernuclear densities. Neutron star matter is thus commonly approximated 
as having zero temperature, so that the equilibrium composition and other 
thermodynamic properties depend on density alone. 
Solving the equilibrium compositions for a given equation of state (EOS) 
at various baryon densities yields the energy density and pressure which 
enable the calculation of global neutron star properties.  

%=================================================================
%    Hyperon Formation Neutron Stars
%=================================================================

\section{Hyperon Formation in Neutron Stars}

In this section we review the principal results of recent studies regarding 
hyperon formation in neutron stars. The masses, along with the strangeness 
and isospin, of nucleons and hyperons are given in
Tab.~\ref{tab:baryons}. The electric charge and isospin combine in determining
the exact conditions for each hyperon species to appear in the matter. Since 
nuclear matter has an excess of positive charge and negative isospin, negative 
charge and positive isospin are favorable along with a lower mass for hyperon 
formation, and it is generally a combination of the three that determines the
baryon density at which each hyperon species appears. A quantitative 
examination requires, of course, modeling of high density interactions. 
We begin with a brief discussion of the current experimental and theoretical 
basis used in recent studies that have examined hyperon formation in neutron 
stars.

\subsection{\it Experimental and Theoretical Background}

The properties of high density matter chiefly depend on the nature of the 
strong interactions. Quantitative analysis of the composition and physical 
state of neutron star matter are currently 
complicated by the large uncertainties regarding strong interactions, both in 
terms of the difficulties in their theoretical description and from the 
limited relevant experimental data. None the less, progress in both 
experiment and theory have provided the basis for several recent studies of 
the composition of high density matter, and in particular suggests it will 
include various hyperon species.

Experimental data from nuclei set some constraints on various physical 
quantities of nuclear matter at the nuclear saturation density, 
$\rho_0\!=\!0.16$ fm$^{-3}$. Important quantities are the bulk binding energy, 
the symmetry energy of non-symmetric matter (i.e., different numbers of 
neutrons and protons), the nucleon effective mass in a nuclear medium, and a 
reasonable constraint on the compression modulus
of symmetric nuclear matter. However, at present, little can be deduced 
regarding properties of matter at higher densities. Heavy ion collisions 
have been able to provide some information regarding higher density nuclear 
matter, but the extrapolation of these experiments to neutron star matter 
is questionable since they deal with hot non-equilibrated matter. 

Relevant data for hyperon-nucleon and hyperon-hyperon interactions is more 
scarce, and relies mainly on hypernuclei experiments (for a review of 
hypernuclei experiments, see Chrien \& Dover (1989), 
Gibson \& Hungerford (1995)). 
In these experiments a single hyperon is formed in a nucleus, and its binding 
energy is deduced from the energetics of the reaction
(typically meson scattering such as $X(K^-,\pi^-)X$). 

There exists a large body of data for single $\Lambda$-hypernuclei, which 
clearly shows bound states of a $\Lambda$ hyperon in a nuclear medium. 
Millener, Dover \& Gal (1988) used the nuclear mass dependence 
of $\Lambda$ levels in hypernuclei to derive the 
density dependence of the binding energy of a $\Lambda$ hyperon in nuclear 
matter.  In particular, they estimate the potential depth of a $\Lambda$ 
hyperon in nuclear matter at density $\rho_0$ to be about $-28$ MeV, which is 
about one third of the equivalent value for a nucleon in symmetric nuclear 
matter. The data from $\Sigma$-hypernuclei are more problematic (see below). 
A few emulsion events that have been attributed to $\Xi$-hypernuclei seem
to suggest an attractive $\Xi$ potential in a nuclear medium, somewhat 
weaker than the $\Lambda-$nuclear matter potential.

A few measured events have been attributed to the formation of double $\Lambda$
hypernuclei, where two $\Lambda$'s have been captured in a single nucleus.
The decay of these hypernuclei suggests an attractive $\Lambda\!-\!\Lambda$ 
interaction potential of $4\!-\!5$ MeV (\cite{BodUs}), somewhat less than the 
corresponding nucleon-nucleon value of $6\!-\!7$ MeV. This value of the 
$\Lambda\!-\!\Lambda$ interaction
is often used as the baseline for assuming a common hyperon-hyperon potential, 
corresponding to a well depth for a single hyperon in isospin-symmetric 
hyperon matter of -40 MeV. While this value should be taken with a large
uncertainty, the typical results regarding hyperon formation in neutron stars
are generally insensitive to the exact choice for the hyperon-hyperon 
interaction, as discussed below.

We emphasize again that the experimental data is far from comprehensive, and 
great uncertainties still remain in the modeling of baryonic interactions. 
This is especially true regarding densities greater than $\rho_0$, where 
the importance of many body forces increases. Three body interactions 
are used in some nuclear matter models (Wiringa, Fiks \& Fabrocini 1988, 
Akmal, Pandharipande \& Ravenhall 1998). Many-body 
forces for hyperons are currently difficult to constrain from experiment 
(\cite{BodUs3bd}), although some attempts have been made on the basis 
of light hypernuclei (\cite{GibHun95}). Indeed, field theoretical models 
include a repulsive component in the two-body interactions through the 
exchange of vector mesons, rather than introduce explicit many body terms. 
We note that the effective equation used here is also compatible with 
theoretical estimates of $\Lambda$NN forces through the repulsive terms it 
includes (\cite{MDG}).

In spite of these significant uncertainties, the qualitative conclusion that 
can be drawn from hypernuclei is that hyperon-related interactions are similar 
both in character and in order of magnitude to the nucleon-nucleon 
interactions. Thus nuclear matter models can be reasonably generalized to 
include hyperons as well. In recent years this has been performed 
mainly with relativistic theoretical field models, where the meson fields are 
explicitly included in an effective Lagrangian. A commonly used approximation 
is the relativistic mean field (RMF) model following Serot \& Walecka (1980), 
and implemented first for multi-species matter by Glendenning (1985), and 
more recently by Knorren et al.\ (1995) and Schaffner \& Mishustin (1996) 
(see the recent review by Glendenning (1996)). A related approach is the
relativistic Hartree-Fock (RHF) method that is solved with relativistic 
Green's functions (\cite{WWH}, \cite{HetalH}). 
Balberg \& Gal (1997) demonstrated that the quantitative results of 
field theoretical calculations can be reproduced by an effective potential
model. 

The results of these works provide a wide consensus regarding the 
principal features of hyperon formation in neutron star matter. This 
consensus is a direct consequence of incorporating experimental data 
on hypernuclei (\cite{BGH}). These principal features are discussed below.

\subsection{\it Estimates for Hyperon Formation in Neutron Stars}

\begin{figure}[htb]
\begin{center}
\figcaption{Relative fractions of the equilibrium composition of neutron star 
matter as a function of the baryon density, for EOS 2: (a) nuclear matter 
(b) matter with nucleons and all hyperons (c) matter with nucleons, $\Lambda$ 
and $\Xi$ hyperons but no $\Sigma$'s. \label{fig:Comps}}
\vspace{2.0cm}
\includegraphics[width=10cm]{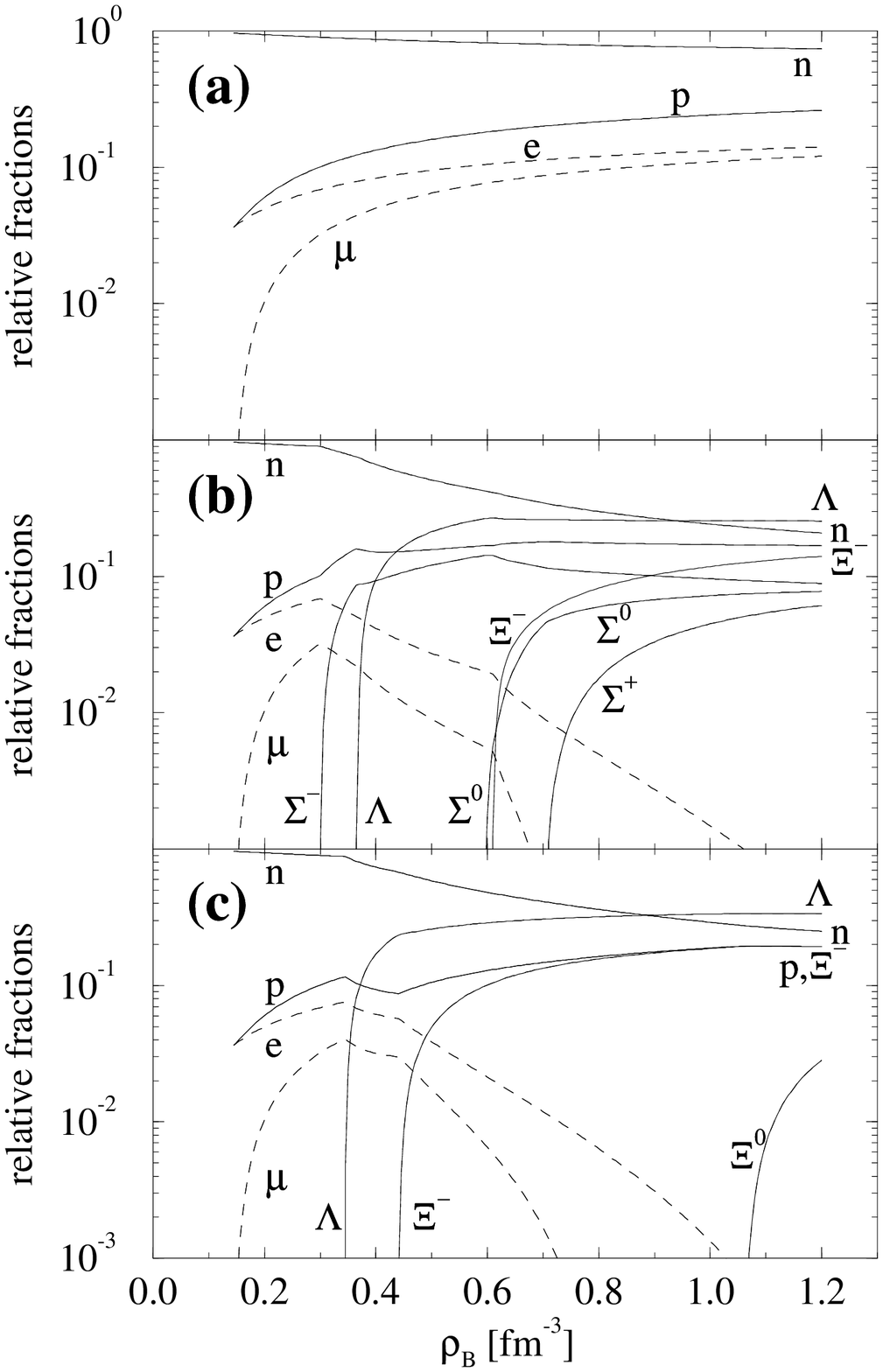}
\end{center}
\end{figure}

Hyperons can form in neutron star cores when the nucleon chemical 
potentials grow large enough to compensate for the mass differences between 
nucleons and hyperons, while the threshold for the appearance of the 
hyperons is tuned by their interactions.
The general trend in recent studies of neutron star matter
is that hyperons begin to appear at a density of about 
$\rho_B\!=\!2\rho_0$, and that by $\rho_B\!\approx\!3\rho_0$ hyperons 
sustain a significant fraction of the total baryon population. An example of 
the estimates for hyperon formation in neutron star matter, as found in many 
works, is displayed in Fig.~\ref{fig:Comps}. The equilibrium compositions 
- relative particle fractions $x_i$ - are plotted as a function 
of the baryon density, $\rho_B$. These compositions were calculated with 
case 2 of the effective equation of state detailed in the appendix, which is 
similar to model $\delta\!=\!\gamma\!=\!\frac{5}{3}$ of Balberg \& Gal (1997). 
Figure~\ref{fig:Comps}a presents the equilibrium compositions for the 
``classic'' case of nuclear matter, when hyperons are ignored, and matter is 
composed of nucleons and leptons. The equilibrium compositions when hyperons 
are included are shown in Fig.~\ref{fig:Comps}b, when the interaction of 
$\Sigma$ hyperons in nuclear matter (nm) is set to be equal to the 
$\Lambda-$nm case (except for the inclusion of isospin dependent components in 
the $\Sigma-$nm case). Key qualitative aspects of hyperon formation in neutron 
star matter are:

\begin{enumerate}  
\item The first hyperon species that appears is the $\Sigma^-$, closely 
followed by the $\Lambda$. The negative charge of the  $\Sigma^-$ outweighs 
the 80 MeV mass difference, as a result of the more lenient condition of
Eq.~(\ref{eq:genequi}) that requires $\mu_\Lambda\!=\!\mu_n$ but 
$\mu_{\Sigma^-}\!=\!\mu_n+\mu_e$. However, the formation of $\Sigma^-$ 
hyperons is quickly moderated by the isospin dependent forces
that disfavor an excess of $\Sigma^-$'s over $\Sigma^+$'s, and also joint 
excess of $\Sigma^-$'s and neutrons (both of negative isospin projection).
Thus, the $\Sigma^-$ fraction saturates 
at about 0.1, while the $\Lambda$'s, free of isospin-dependent forces, 
continue to accumulate until short range repulsion forces cause them to
saturate as well.
\setlength{\parskip}{0.0in}
\item Other hyperon species follow at higher densities. Under the assumptions 
of the particular model of EOS 2, other $\Sigma$'s generally 
appear before the $\Xi$'s due to the large mass difference, 
but the $\Xi^-$ becomes favored due to its negative 
electric charge and quickly becomes abundant in the matter. 
\setlength{\parskip}{0.0in}
\item A unique aspect of hyperon accumulation is the immediate 
deleptonization of the matter. Leptons are rather expensive in terms of energy 
density (and pressure), and survive in nuclear matter only in order to 
maintain charge neutrality with the protons. 
Hyperons offer an option for lowering the neutron excess free of lepton 
formation, and the negatively charged hyperons allow charge neutrality to 
be maintained within the baryon community. The lepton fraction is therefore reduced 
by hyperon formation, and the appearance of the $\Xi^-$ is followed by a very 
powerful deleptonization. The muon population is completely extinguished, 
and the electron fraction drops below $1\%$, whereas it exceeds $10\%$ in the 
nuclear matter case. 
\end{enumerate}

We remark that some of these general features are somewhat dependent on the 
assumptions used to describe the hyperon-nucleon interactions. In particular, 
if any reaction is changed to be highly repulsive, the formation of some 
species may become suppressed. As an example, Fig.~\ref{fig:Comps}c 
shows the equilibrium compositions found when a strongly 
repulsive component is introduced in the potential of $\Sigma$ hyperons 
in nuclear matter. The existence of such a repulsive isoscalar component 
has been suggested on the basis of recent analysis of $\Sigma^-$ atoms
(\cite{SigAtomB,SigAtomM}). The analysis predicts a ($\Sigma-$nm) repulsion 
of several MeV at the nuclear saturation density, and even larger repulsion 
at greater densities. If such repulsion exists, $\Sigma$'s 
do not form in neutron star matter (see also \cite{SBH,BGH}). 
As a result, $\Lambda$ formation begins at slightly 
lower densities than when $\Sigma$'s are present, and $\Xi$ formation is 
especially enhanced. It is noteworthy, however, that the overall strangeness 
fraction in this case is similar to the case when $\Sigma$'s are present. 
Since at least the $\Lambda$-nm interaction seems well determined, we believe 
a significant change of the basic features of hyperon formation is 
unlikely (see also the analysis by Glendenning \& Moszkowski (1991)). There 
is less dependence on the hyperon-hyperon interactions (again, unless they are 
set to be highly repulsive - which seems unlikely in view of data from double 
$\Lambda$ hypernuclei). This is because the matter is dominated by nucleons 
until high densities, where universal short range forces are expected to 
take precedence over the specific baryon identities.

We note in passing that $\Delta$ isobars are also candidates for formation 
in high density matter. Most works which examined the possible appearance 
of $\Delta$ isobars in dense matter find that they are never present, due 
to a strong isovector repulsion. It should be noted
that in some relativistic Hartree-Fock frameworks the nucleon-$\Delta$ 
coupling (through the $\rho$-meson) is significantly weakened, and $\Delta$
isobars are found to appear in high density neutron star matter 
(\cite{WWH}, \cite{HetalH}). In this work we follow the assumption that 
$\Delta$'s do not appear in neutron star matter. 

To conclude, it is noteworthy that recent works agree that hyperons appear 
at a density of about $2\rho_0$, and at higher densities the matter will 
possess a sizable hyperon fraction, coupled to  significant deleptonization. 
We emphasize again that these qualitative features are common to all works 
which examined hyperon formation in neutron star matter and are only weakly 
dependent on the specifics of the underlying models. These models are based 
on various types of approximations and are limited by the large uncertainties 
involved; clearly, further work (and, hopefully, more experimental data) is 
required to obtain more reliable quantitative results. None the less, this 
consensus is a direct consequence of employing data from hypernuclei 
experiments, and therefore may serve as valid indication regarding 
hyperon abundances in high density matter at beta-equilibrium. In the 
following analysis we assume that these recent works do provide a basis 
for the investigation of the effects of hyperon formation on global 
properties of neutron stars.
  
%=================================================================
%    Equation of State
%=================================================================

\section{Roles of Hyperons in the Equation of State}

The most significant implications of the composition of high density matter 
for the global properties of neutron stars are reflected 
in the equation of state (EOS). It is the EOS which determines
the mass and radius a neutron star can hold for a given central density, and 
what effect rotation will have on these values. In turn, the 
observed constraints on the maximum mass and rotational frequencies of pulsars 
can provide indirect clues regarding the physics of high density matter.

The principal effect caused by hyperon formation in the dense core 
of neutron stars is a softening of the EOS. The softening 
is seen when compared against the EOS for matter composed of nucleons 
and leptons alone, but with otherwise identical assumptions regarding 
the strong interactions. This basic property of matter with hyperons has 
been noted in many works (\cite{Glenbook}, and references therein), 
and is a fundamental result that is basically 
independent of the precise model used for the baryonic interactions. 
Hyperons offer another degree of freedom for baryonic matter, 
and relieve some of the Fermi pressure of the nucleons. The creation 
of additional species allows energy to be held as mass rather than 
kinetic and potential energy, which are more expensive 
in terms of pressure. Here we wish to call attention to some 
of the underlying features of hyperon induced softening.

It is useful to begin with an emphasis on the fundamental relation 
between the microscopic baryonic interactions and the macroscopic EOS. 
As demonstrated explicitly  by Pandharipande, Pines \& Smith (1976), 
the nuclear matter EOS is critically coupled to the nucleon-nucleon 
interaction: the greater the short range repulsion, the greater the 
energy density for a given baryon density, and hence a stiffer EOS. 
Since there are practically no experimental limits on the short range 
repulsion, published nuclear matter equations vary over a 
relatively large range. 

The microscopic-macroscopic connection becomes even more pronounced when 
hyperons are taken into account. As discussed in detail by Balberg \& Gal 
(1997), the nucleon-nucleon interaction also determines the rate at which 
the nucleon chemical potential rises with increasing density. Since the 
nucleon chemical potentials determine the amount of hyperon accumulation 
in the matter according to the conditions of Eq.~(\ref{eq:genequi}), 
strong short range nucleon-nucleon repulsion enhances hyperon formation, 
which has a softening effect on the EOS. 
Hence, hyperon formation induces a fundamental balance 
between the microscopic equilibrium compositions and the macroscopic 
properties of the EOS, which restrains the resulting equation to a 
relatively narrow range of values. {\it Hyperon formation therefore 
serves as a ``pressure control'' mechanism} in high density matter.

We demonstrate the ``pressure control'' induced by hyperons by comparing two 
equations of state which are significantly  different in their description 
of baryon-baryon interactions. Both are based on the effective EOS of Balberg 
\& Gal (1997) (see appendix). EOS 1 is moderately stiff, and has an 
incompressibility of $K\!=\!240$ MeV for symmetric nuclear matter at density 
$\rho_0$, which is the commonly used value. EOS 2 is stiff, with $K\!=\!320$ 
MeV. The calculated equations of state for matter in beta-equilibrium are 
plotted in Fig.~\ref{fig:EOSS}; plotted are the EOS for (i) matter with 
nucleons and leptons alone, (ii) matter with nucleons, hyperons and leptons, 
and (iii) matter with nucleons, hyperons and leptons, but when $\Sigma$ 
hyperons are absent from the matter (i.e., a strongly repulsive $\Sigma$-nm 
interaction is assumed). 

The pressure control discussed above can be understood from the qualitative
difference between the respective equations of model 1 and model 2. In the 
nuclear matter case (top solid line and dashed line), EOS 2 is stiffer than 
EOS 1 through the entire density range since it is based on a more powerful 
short range repulsion between nucleons. No further degrees of freedom exist, 
and the difference between the two equations grows unhindered. A qualitatively 
different picture arises when comparing the two equations if all hyperon 
species are allowed to appear (thick solid curve and thick dashed curve for 
EOS 1 and 2 respectively). The stronger repulsion between nucleons in EOS 2 
enhances hyperon accumulation, resulting in a more pronounced softening effect.
Hyperon formation actually causes EOS 2 to become softer over some 
density interval, and in general confines both equations to a narrower range 
of values.

It is also noteworthy that a similar pressure control is achieved when 
$\Sigma$ hyperons are extinct (dot-dashed lines). Suppressing $\Sigma$ 
formation eliminates some degrees of freedom and the resulting EOS is
naturally stiffer than when all hyperon species appear, 
implying that the EOS depends on the number of available species 
(\cite{SBH}, \cite{BGH}). 
None the less, the remaining hyperons maintain enough degrees of freedom to 
allow for the manifestation of the basic feature of pressure control. 
 
Similar trends of pressure control are found for other variations
of the effective EOS (\cite{BGH}), and may also be inferred 
from other works which compared equations of state 
(\cite{GH}, \cite{SchaabAA}). This is yet another consequence of the common 
basis used for the hyperon-nucleon interactions, and the same reservations 
made about hyperon formation apply here as well: if the hyperon-nucleon 
interactions at higher densities are radically different than those inferred 
from hypernuclei, the pressure regulation effect could be lost.

In conclusion, we emphasize that ``pressure control'' is a fundamental result 
of the availability of new baryonic degrees of freedom. Since the 
bulk of the matter in neutron stars is at a density close to that of the 
core, the softening of the EOS and the hyperon induced pressure control 
have immediate consequences on the global properties of these stars.
These consequences are discussed in the following subsections.

\subsection{\it The Maximum Mass}

The most fundamental role played by the high density EOS is in determining
the relation between the star's central density and its mass and radius.
Calculating the mass as a function of 
the central density yields a mass sequence for a given 
EOS and these sequences provide a convenient measure for
comparing different equations. Of special importance is the maximum mass 
each equation predicts, since it serves as an integral measurement of the 
properties of the equation.    

We begin with an examination of static (non-rotating) sequences for the 
different equations of state, by integrating the Tolman-Oppenheimer-Volkoff 
equations, namely
\begin{equation} \label{eq:TOV}
\frac{dm}{dr}=4\pi r^2\varepsilon\;\;  ; \;\;
\frac{dP}{dr}=-\frac{Gm\varepsilon}{r^2}
               \left(1+\frac{P}{c^2\varepsilon}\right)
               \left(1+\frac{4\pi r^3P}{c^2m}\right)
                       \left(1-\frac{2Gm}{c^2r}\right)^{-1} \;\;,
\end{equation}
where $\varepsilon$ is the energy density (in gm/cm$^3$) and $P$ is the 
pressure (in dyne/cm$^2$). The integration is performed 
following the recipe of Arnett and Bowers (1977). For subnuclear 
densities we use the EOS of Feynman, Metropolis \& Teller 
(1949), followed by that of Baym, Pethick \& Sutherland (1971) up to the 
neutron drip density, and the equation of Baym, Bethe \& Pethick (1971) up 
to nuclear matter density. We typically interpolate over a small region 
when connecting different equations (while the choice of interpolation 
limits is somewhat arbitrary, the resulting neutron star properties usually 
show very low sensitivity to these limits).

\begin{figure}[htb]
\begin{center}
\figcaption{Equations of State for model 1 (thick lines) and model 2 (thin 
lines). The equations correspond to nuclear matter (solid lines), matter 
with nucleons and all hyperons (dashed lines), and matter with nucleons, 
$\Lambda$ and $\Xi$ hyperons but no $\Sigma$'s. (dot-dashed lines)
\label{fig:EOSS}}
\vspace{0.0cm}
\includegraphics[width=10cm]{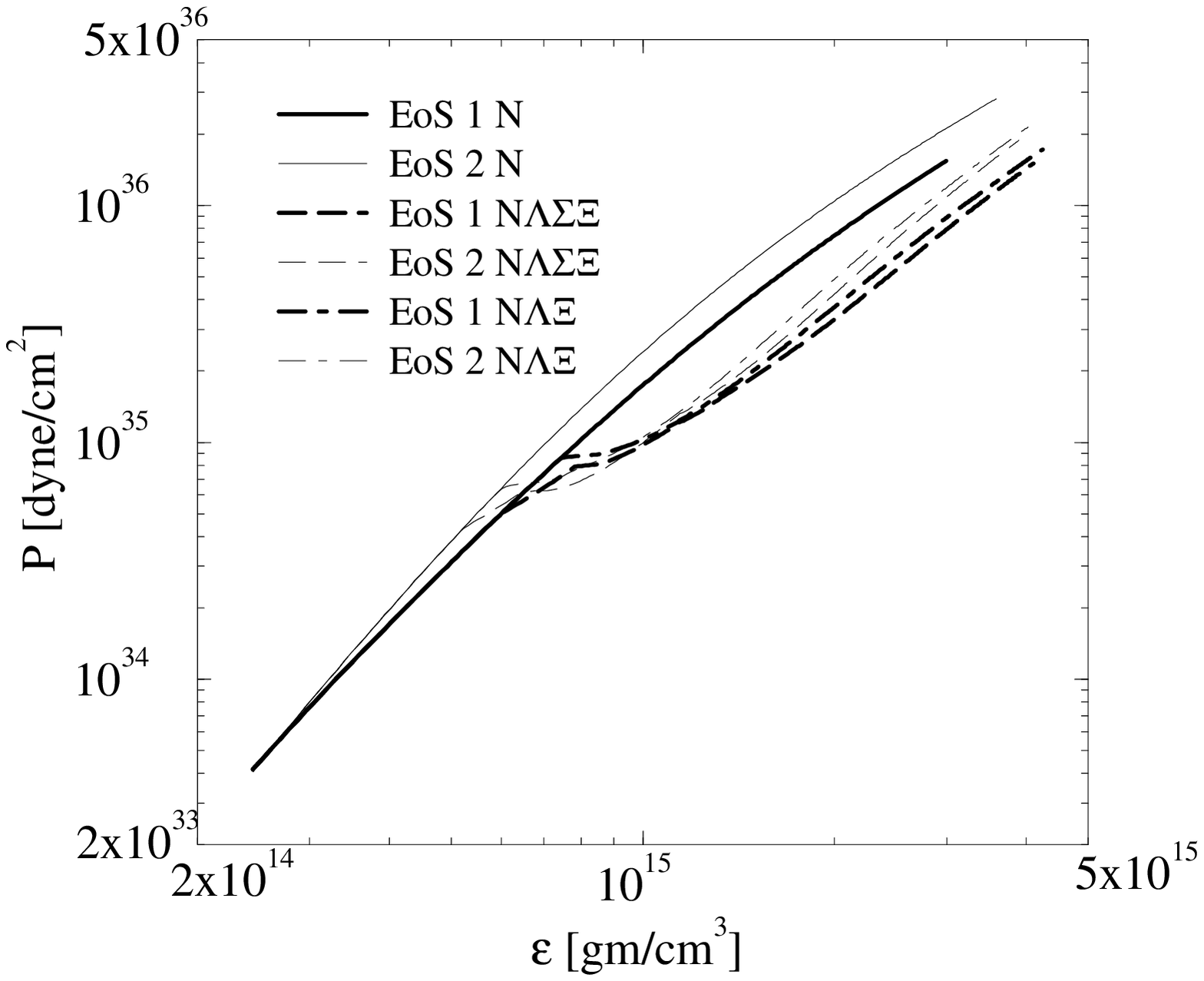}
\vspace{-4.0cm}
\figcaption{Static neutron star masses (in units of $M_\odot$) as a function 
of the central energy density, $\varepsilon_c$, for the equations of state 
presented in Fig.~\ref{fig:EOSS}. All lines as indicated in
Fig.~\ref{fig:EOSS}. \label{fig:M_vs_epsc}}
\vspace{0.0cm}
\includegraphics[width=10cm]{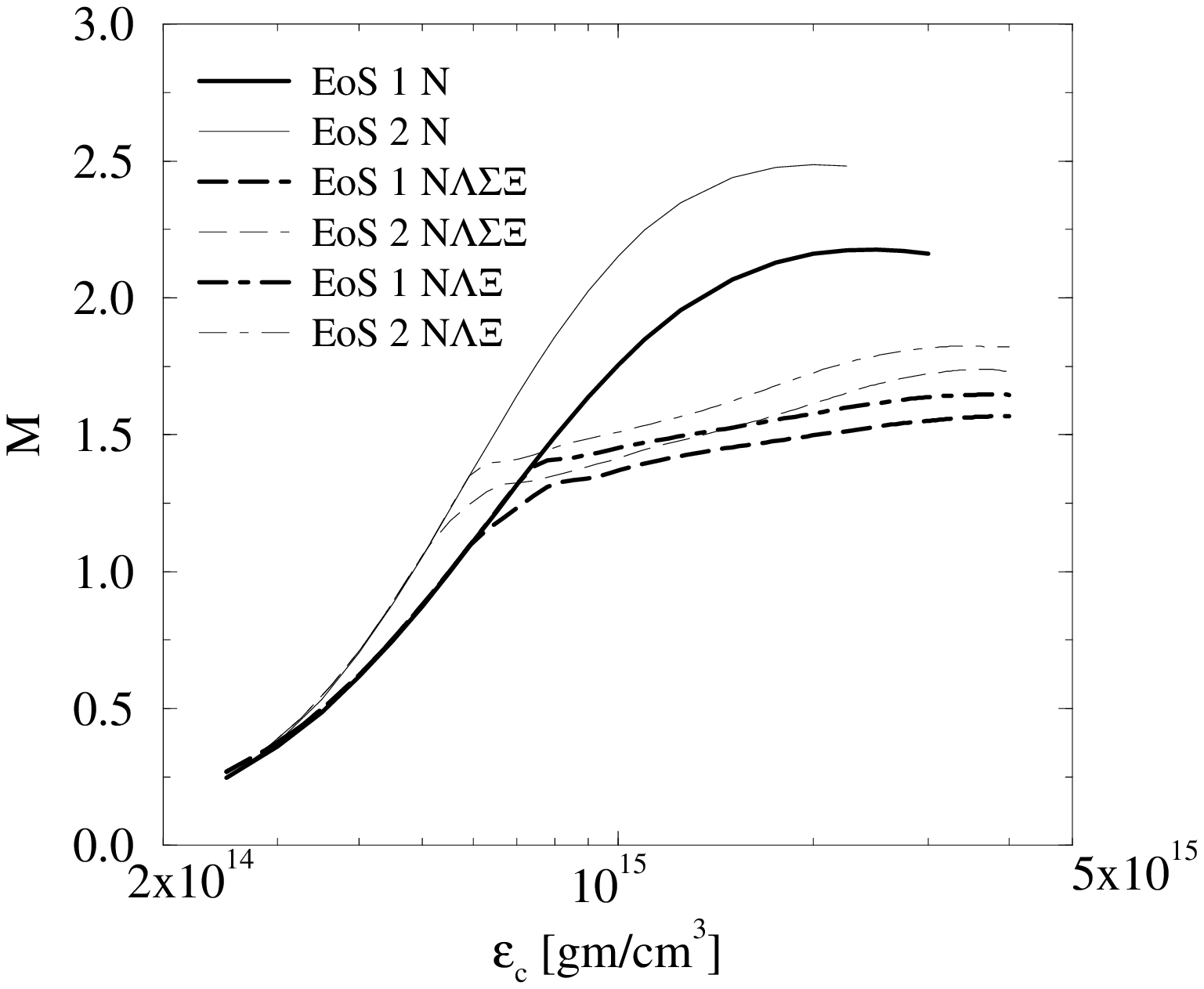}
\end{center}
\end{figure}

The resulting mass sequences are shown in Fig.~\ref{fig:M_vs_epsc}. The onset 
of hyperon formation can be identified clearly for every EOS as the 
irregularities in each curve. The softening induced by the formation of 
hyperons is also easily identified, since it enforces - for any given central 
density - a lower neutron star mass than the mass found for nuclear matter 
with an equivalent EOS. 

Correspondingly, the maximum mass found with an EOS which includes hyperons 
is naturally lower than when hyperons are neglected. Again, this is a general
result of allowing more baryon species to appear in the matter, regardless
of the specifics of the model used for the strong interactions (and, indeed, 
it is noted in all works which included hyperon formation in neutron 
star matter). A more subtle effect concerning neutron star masses is 
the theoretical limit that the inclusion of hyperons forces on the range 
of values of the maximum mass. This can also be seen in
Fig.~\ref{fig:M_vs_epsc}, where the 
different equations with hyperons yield maximum masses which lie in a rather
narrow range: $1.5-1.8\;M_\odot$. 

Limiting the range for the maximum mass is not unique to the specific 
equations used in this work. As was first demonstrated in the framework 
of RMF models (\cite{GlenMozs91}), this restriction basically arises from 
constraining the hyperon-related interactions by hypernuclei experimental 
data.  We emphasize here that the underlying principal reason for this
maximum mass constraint is the hyperon-induced pressure control 
discussed above, and therefore is, in essence, model independent.
This conclusion is further supported by a survey of maximum masses found 
in various works where hyperons were included in high density matter: 
a large majority of these works (see the reviews by Glendenning 1996 and 
Prakash et al.\ 1997) place the maximum static mass in a narrow range of
$1.5\!-\!2.0\; M_\odot$, with the upper limit being reached only with equations 
that are extremely stiff at $\rho\!\approx\!\rho_0$. Since no ``pressure 
control'' is available for nuclear matter equations of state, the 
theoretical limit they provide on the static maximum mass is much weaker: 
roughly $1.5\!-\!2.7\;M_\odot$ (Cook, Shapiro \& Teukolsky 1994).

Unfortunately, the current observational constraint is only that 
$M_{max}\!\geq\!1.44\;M_\odot$ (the well determined mass of pulsar 1913+16).
This constraint allows almost all theoretical 
equations of state to be considered legitimate. 
The fact that larger mass pulsars have not been observed may indicate
that the maximum mass is indeed low, and several arguments have 
been made in support of this possibility (\cite{BB}). On the other hand, should 
a large mass neutron star be observed, it will prove
extremely valuable in ruling out different equations of state. 
Currently, the Vela pulsar is the only likely candidate for a large mass 
pulsar (Van Kerkwijk, M. 1997, private communication), but the uncertainties
in determining its mass are still very large. It should be noted 
that recently measured kilohertz Quasi-Periodic-Oscillations (QPO's) in X-ray 
binaries (\cite{QPO}) may also provide tighter constraints on 
the value of the maximum mass, since the underlying neutron star is known to 
be accreting.

\subsection{\it Radii}

Solutions of the Tolman-Oppenheimer-Volkoff equations relate the radius of 
a neutron star to its mass for any given EOS. Broadly speaking, the radius of 
the neutron star is a poor indicator of the properties of the inner core 
($\rho_B\!\geq\!2\rho_0$). While this inner core holds most of the mass of 
a $\sim\!1.4\; M_\odot$ star for almost any EOS, it only extends to about 
half of the star's radius.
The radius is far more dependent on the EOS at 
$\rho_B\!\sim\!\rho_0$ and below, which is indifferent 
to the possible appearance of new degrees of freedom in the core.
Never the less, 
some qualitative observations regarding possible effects of hyperon formation 
on neutron star radii can still be made.

\begin{figure}[htb]
\begin{center}
\figcaption{Radius vs. gravitational mass (in units of $M_\odot$) 
relations for static neutron stars calculated with EOS 1 (thick solid line) 
and EOS 2 (thin solid line), and for the nuclear matter equations FPS, A, AU 
and L (see text for details).
\label{fig:M_vs_R}}
\vspace{0.0cm}
\includegraphics[width=10cm]{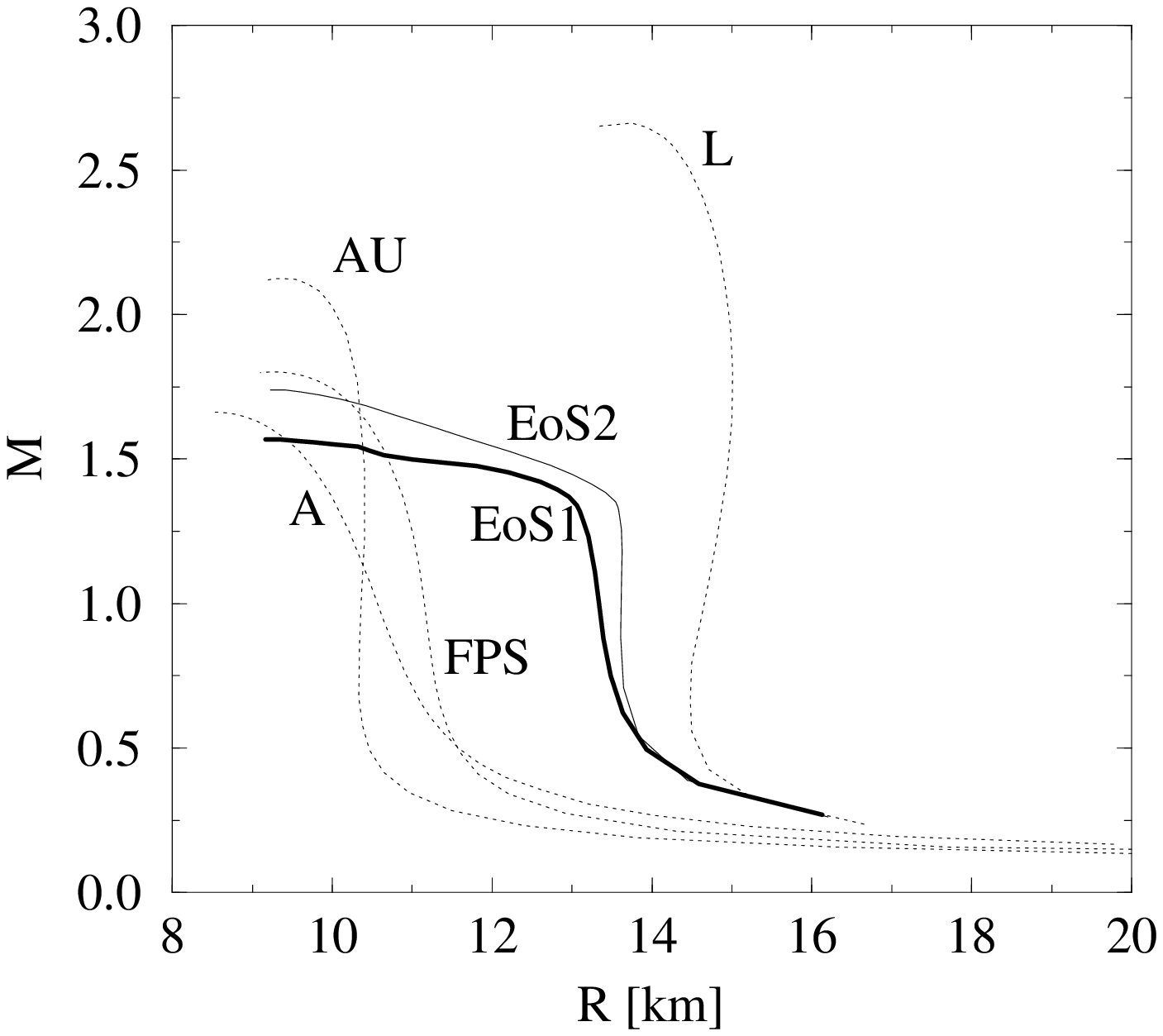}
\vspace{-5.0cm}
\figcaption{Constant rest mass sequences for EOS 2 showing the angular 
velocity, $\Omega$, as a function of the angular momentum, $J$.
Selected sequences are labeled by the value of the rest mass, and the sequence
which has a static gravitational mass of $1.4\;M_\odot$ is marked with an 
asterisk. The mass shed limit is the bold dashed line and the quasi-radial 
stability limit is denoted by the thin dashed line. The inset shows an 
expanded view of the region near the maximum mass model (open circle) and 
shows the location of the maximum $\Omega$ model located at the intersection 
of the mass shed and stability limit.
\label{fig:Ohm_vs_J}}
\vspace{0.0cm}
\includegraphics[width=8cm]{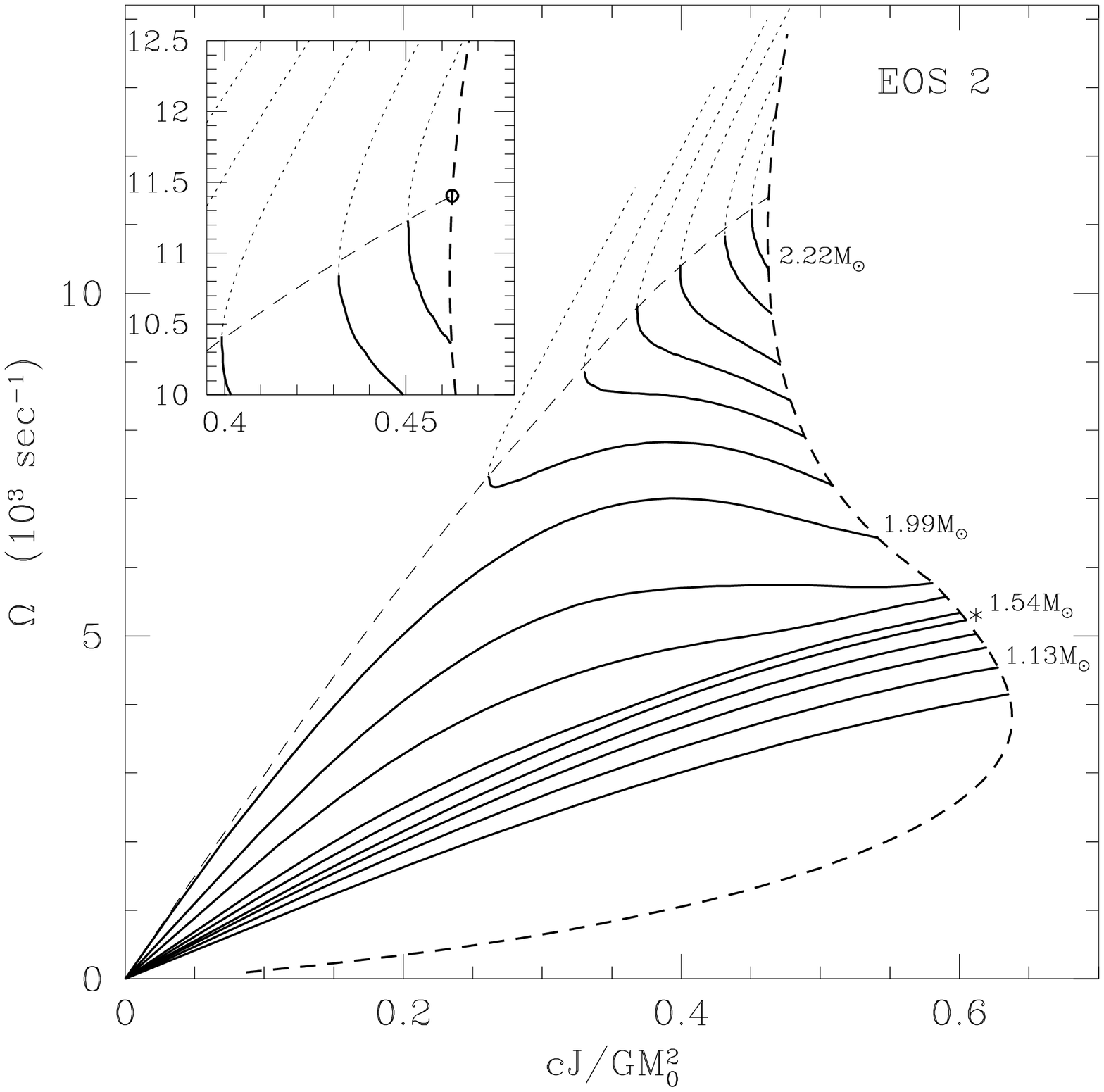}
\end{center}
\end{figure}

Figure~\ref{fig:M_vs_R} compares the mass-radius dependence of static neutron 
stars for various published equations of state which do not include hyperons 
to those of EOS 1 and 2 with all types of hyperons. The nuclear matter 
equations are FPS (Lorenz, Ravenhall \& Pethick 1993), L (Mean-field EOS by 
Pandharipande \& Smith 1975), A (Reid soft-core by Pandharipande 1971b), and 
AU (Wiringa, Fiks \& Fabrocini 1988), where we follow the notation of Cook et 
al.\ (1994). The radii at low masses ($M\!\leq\!1\;M_\odot$) are 
basically dependent only on the EOS below $2\rho_0$, and the differences 
between the radii for various equations of state in this region reflect the 
differences regarding nuclear matter at these densities. For masses close to 
the maximum mass for each equation, the radius is naturally smaller and the 
star is typically more compact when the high density EOS is softer.

One effect that does stand out concerning the equations with hyperons is that 
there is a larger difference between the typical radii for a low mass 
star and the radius of the maximum mass configuration than in other equations. 
Comparing the radius of a $1.4\; M_\odot$ star and the radius for the maximum 
mass star yields a difference of 3.5 km and 3.4 km for EOS 1 and 2, 
respectively. For the other equations this difference does not exceed 2.5 km. 
This is a result of the specific contrast in the equations 
with hyperons used here, which are moderately stiff to stiff at lower 
densities, and soft (due the effect of hyperons) at higher densities. Clearly, 
such an effect will be common to any EOS which follows such a change from 
low to high densities; however, hyperon formation offers a natural 
explanation for such a trend, if it indeed exists. It should also be noted 
that a large difference between the radii at $1.4\;M_\odot$ and at maximum 
mass is not a necessary consequence of hyperon formation, and would not have 
been found if the EOS was softer at low densities.         

Unfortunately, current observations of neutron stars do not provide radius 
measurements to any useful precision. There is no reason to infer either 
presence or absence of a large difference between the radius for 
$M\!=\!1.4\!\;M_\odot$ and $M\!\la\!M_{max}$, which is not known in 
any case (however, as discussed below, interpretation 
of pulsar glitches may serve as an indication that the equation of state 
does change from stiff at low densities to softer at higher ones).
We note that future analysis of QPO's might hold significant potential for 
establishing mass radius relationships for accreting objects, although 
accurate measurement of the neutron star rotation and its effect on the 
stellar shape are required (Miller, Lamb \& Cook 1998).

\subsection{\it Rotation Periods and Limits}

Constraints on the high density EOS can be derived from the maximal observed 
angular velocity of pulsars, from the properties of their rotational 
evolution (spin down), and from mass-angular velocity relations that have been 
established for a few pulsars. Very rapid rotation 
($\Omega\!\geq\!3\!\times10^4\;$s$^{-1}$), if 
observed, will serve as an important component in determining the structure 
of a neutron star (\cite{SBrapid}). Note that rotational limits and 
deformation must be treated self-consistently in the framework of general 
relativity.

In Fig.~\ref{fig:Ohm_vs_J} we show the dependence of the angular velocity, 
$\Omega$, of a neutron star on its angular momentum, $J$, calculated with the 
formalism presented by Cook et al.\ (1994) with EOS 2. The figure describes 
the star's rotational evolution as it slows down by radiating energy and 
angular momentum. The evolutionary sequence of a ``normal'' star of constant 
rest mass (solid lines) proceeds from the mass shed limit (dashed line on the 
right) to the static limit (the origin) by losing angular momentum. 
As with all equations of state (\cite{CookST94}), there are also 
``supramassive'' sequences, where the rest mass is too large to allow 
a stable static solution. A supramassive sequence is meta-stabilized by 
sufficiently rapid rotation, and will collapse to a black hole at some point 
in its evolution. The onset of collapse (instability to quasi-radial 
perturbations) corresponds to the stability limit, denoted by the thin dashed 
line in the figure.

We call attention to the fact that for EOS 2 there are ``normal'' sequences 
that show spin-up of the neutron star during some part of the sequence. Spin 
up must occur in the super-massive sequence, since their unstable portions
are always at higher angular velocity than the stable portion for the same
value of angular momentum (\cite{CookST94}). On the other hand, spin up of a 
normal sequence is unusual in equations of state for nuclear matter, but is 
possible once hyperons are included. 

\begin{figure}[htb]
\begin{center}
\figcaption{The adiabatic index, $\gamma$, for EOS 2 (solid line), the EOS 
with hyperons of Glendenning (1996) (dashed line) and the nuclear matter 
equation FPS (dot-dashed line), as a function of the mass-energy density.
\label{fig:Gam_vs_eps}}
\vspace{0.0cm}
\includegraphics[width=10cm]{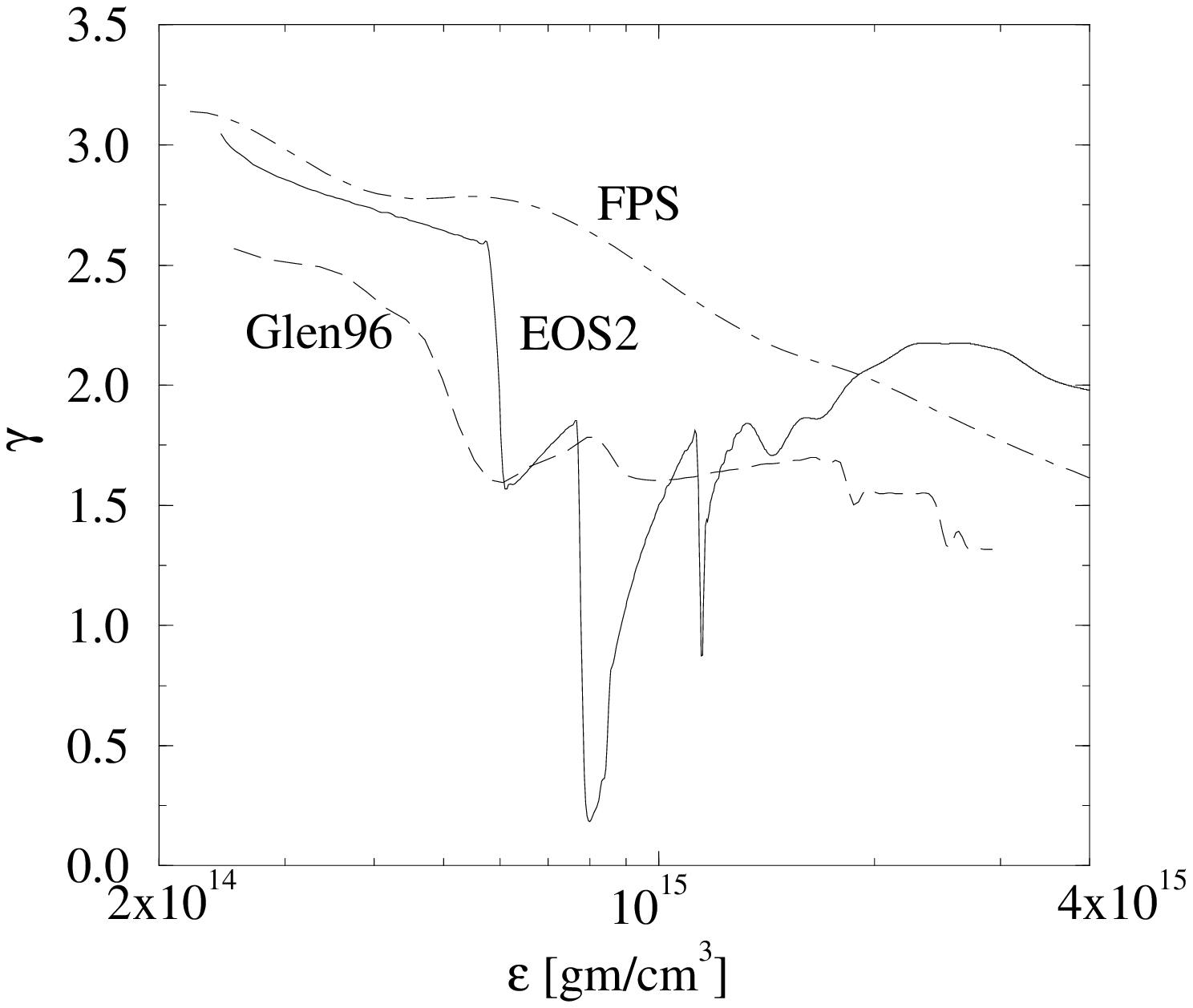}
\vspace{-5.0cm}
\figcaption{Constant rest mass sequences for EOS 2 showing the angular 
velocity, $\Omega$, vs. the gravitational Mass. The filled circles with error 
bars for the mass are the observed values of mass and angular velocity for 
several binary pulsars (see Cook et al.~1994). The vertical long-dashed line is 
the suggested lower limit of $1.55\;M_\odot$ on the mass of Vela X-1. The 
horizontal long-dashed line is the angular velocity of the millisecond pulsar 
PSR 1937+21.
\label{fig:Ohm_vs_M}}
\vspace{0.0cm}
\includegraphics[width=8cm]{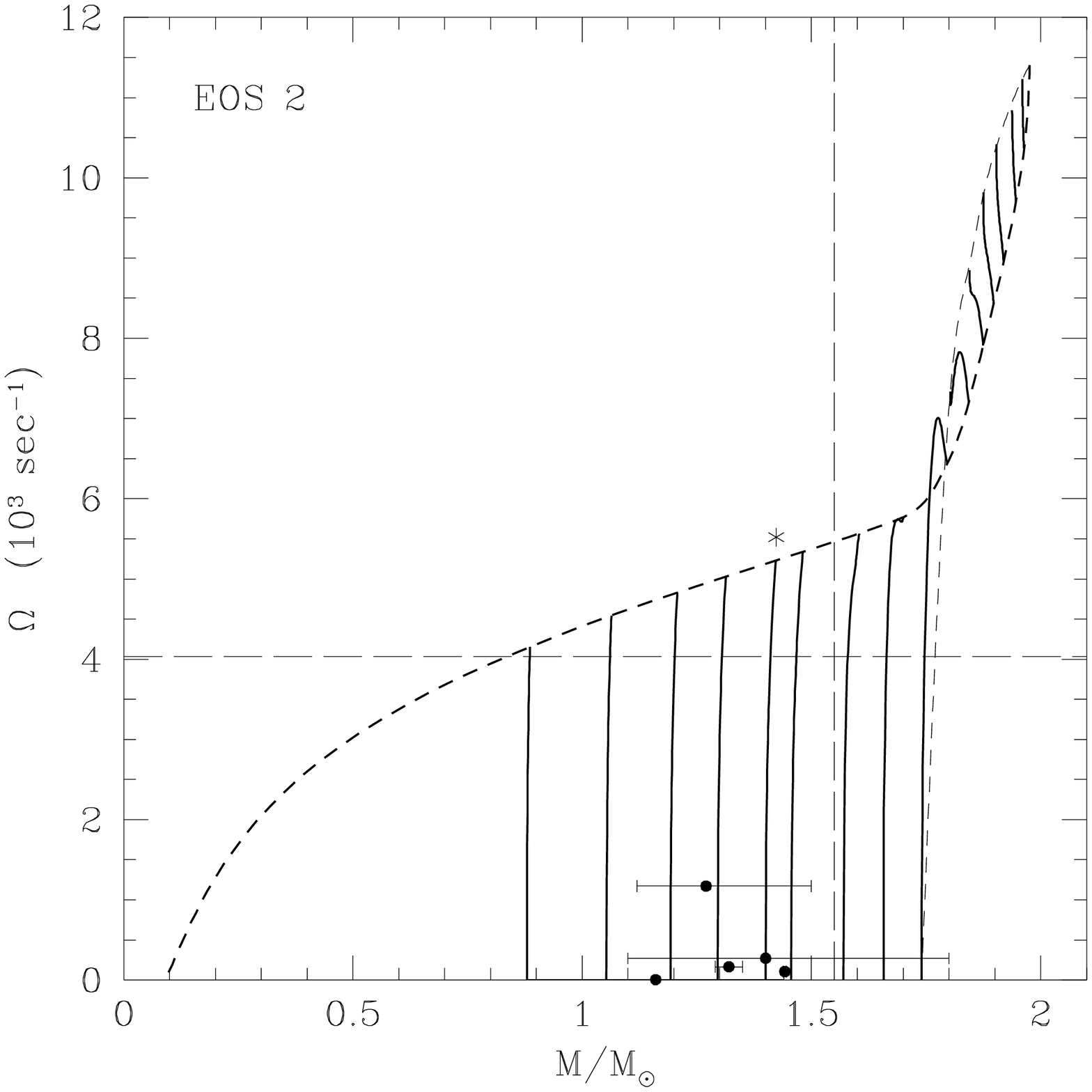}
\end{center}
\end{figure}

Loss of angular momentum causes the star to both lose rotational energy, 
{\it and} contract and become more spherical. The balance of these two 
effects usually results in a decrease of the angular velocity, but an increase 
is also possible if the EOS is sufficiently soft over a large region of the 
star. In Newtonian physics the condition is that the effective adiabatic 
index of the star is less than $\frac{4}{3}$, and when general relativity 
is included, even slightly above this value (Cook, Shapiro \& Teukolsky 1992). 
In Fig.~\ref{fig:Gam_vs_eps} we show the adiabatic index, 
$\gamma\!=\!d\!\log(P)/d\!\log(\rho)$ for EOS 2 and the EOS with hyperons of 
Glendenning (1996, pg. 244). The effect of hyperons on the equations is 
obvious, as $\gamma$ drops considerably at every density that a new species 
appears. For comparison we also show $\gamma$ for the FPS equation, which 
does not include baryon degrees of freedom beyond nucleons. Indeed, spin up 
is not found in any normal sequence of the FPS equation, and also does 
not occur for the EOS with hyperons of Glendenning 
(Glendenning, N. K. 1997, private communication) because the variation in the
adiabatic index is not large enough. The specific details of EOS 2, however, 
lead to an enhanced effect of hyperon formation on the adiabatic index, 
which is why we find that spin up is possible for this equation, even for 
some of the normal neutron star sequences. 

Assuming that equations of state which include only nucleons cannot have 
an effective adiabatic index low enough to allow spin up, a neutron star
that spins up without accreting and which does not collapse may serve as
important indication 
of a more complex structure of the core. Spin up during a pulsar's evolution
should in principle be easy to observe, specifically through the breaking 
index, defined as $n\!\equiv\!\Omega\ddot{\Omega}/(\dot{\Omega})^2$. 
The breaking index of observed pulsars is measured to good precision, and is 
typically found to be 2-3. A pulsar going from spin up to spin down 
should show a breaking index going to $-\infty$ at maximum frequency, and then 
decreases from $+\infty$ as spin down begins. Glendenning, 
Pei \& Weber (1997) suggested that such behavior of the breaking index may 
signal the creation of a mixed baryon-quark phase. It is our purpose here to 
point out that spin up followed by spin down is also a possible (though not 
a necessary) result of hyperon formation. 

Note, however, that spin up occurs for EOS 2 only in high mass stars 
(rest mass larger than $1.87\;M_\odot$) which could be uncommon due to 
selection effects in the pulsar formation mechanism. Furthermore, spin up of
stable sequences is only found for very rapidly rotating configurations where
nonaxisymmetric instabilities (driven by gravitational radiation) may set in. 
This further limits the combinations of mass and rotation period that allow 
spin up in a stable sequence, and could explain why such 
evolution has not been observed, even if physically possible.

We conclude our discussion of rotational properties and limits of neutron stars
in the context of hyperon formation with Fig.~\ref{fig:Ohm_vs_M}, which 
displays the angular velocity versus gravitational mass of constant rest mass 
sequences for EOS 2 with all types of hyperons. Also plotted in the figure
are the known masses and angular velocities for various observed pulsars
(see Cook et al.\ 1994). The horizontal 
dashed line is a minimum-$\Omega$ limit of $\Omega\!=\!4032\;$sec$^{-1}$
set by PSR 1937+21.
The vertical dashed line corresponds to a mass of $1.55\;M_\odot$, which is 
a suggested lower limit for the Vela pulsar mass (still under debate).
 
One finds that current combinations of observed pulsar masses and angular 
velocities do not offer significant constraints on the high density EOS, and, 
in general, are consistent with hyperon formation 
(for which EOS 2 may be taken as representative). Note that the maximum 
angular velocity found for EOS 2 is $\sim 1.15\!\times\!10^4\;$sec$^{-1}$, 
well in the range of values of ``typical'' equations of state. This maximum 
angular velocity is also in good agreement with the ``empirical'' formula of 
Haensel \& Zdunik (1989): 
\begin{equation}
\Omega_{max}=\chi\left(\frac{M_{max}}{M_\odot}\right)^{1/2}
    \left(\frac{R(M_{max})}{10\;\mbox{km}}\right)^{-3/2}\;\mbox{sec}^{-1}
\end{equation}
where $M_{max}$ and $R(M_{max})$ are the gravitational mass and radius of the 
maximum mass static configuration, respectively. The numerical  coefficient 
$\chi$ was found by Cook et al.\ (1994) through a best fit  
to be $\chi\!\approx\!7840\;$sec$^{-1}$, when including the supramassive 
sequences.
In general, the softer the EOS, the larger the predicted maximum angular 
velocity.  Since the mistaken measurement from the remnant of 
SN1987A, there has been much speculation regarding the constraints a 
$0.5\;$msec pulsar, if found, would place on the high density EOS. We 
have not found an EOS where the hyperon induced softening is sufficient 
to allow a $0.5\;$msec pulsar, in agreement with other published works. 

\subsection{\it Crustal Size and Pulsar Glitches}

Pulsar glitch phenomena have been suggested as a probe of neutron star 
properties (Link, Epstein \& Van Riper, 1992). The basic argument is that 
the interpretation of pulsar glitch phenomena suggests a relatively large 
crust, which in turn implies a stiff EOS at supernuclear 
densities, if a pulsar mass greater than 1 $M_\odot$ is assumed. Indeed, 
pulsar glitches are often presented as observational proof that the 
supernuclear density EOS is stiff (\cite{Alparal}). However, a stiff EOS also 
leads to a large value for the maximum mass, which, as discussed 
above, is currently difficult to support by observation. We argue here that 
hyperon formation provides a natural route to combine large crusts 
{\it and} a relatively low maximum mass.   

Glitches are sudden increases in the rotation frequency of pulsars. The 
post-glitch behavior of the pulsar indicates a change in the spin-down rate, 
$\Delta\dot{\Omega}/\dot{\Omega}$, ranging from a fraction of a percent 
(the Crab) to a few percent (the Vela). The generic interpretation 
of glitches suggests a coupling and decoupling process between 
different components of the pulsar (\cite{ShaTeu83}). In this event angular 
momentum is transformed from some weakly coupled component to the bulk of the 
star, which is strongly coupled to the crust through the magnetic field. 
This generic interpretation known as the ``two-component model'', can be 
shown to predict that
\begin{equation} \label{eq:tcmII}
    \frac{\Delta\dot{\Omega}}{\dot{\Omega}}=\frac{I_c}{I_{tot}}\;\;\;,
\end{equation}
where $I_{tot}$ is the total moment of inertia of the pulsar, and $I_c$ is 
the moment of inertia of the more rapidly rotating component.

The most successful model suggested so far for pulsar glitches has been the
vortex creep theory (\cite{PinAl}, \cite{Alparal}).
In this model, the glitches are driven by 
pinning and unpinning of vortices of the neutron superfluid and the lattice
of neutron rich nuclei which coexist in the inner crust of pulsars. Assuming 
that vortex creep can occur between the density of neutron drip 
($\approx 4\times10^{11}$ gm/cm$^3$) to about half the nuclear saturation 
density ($\approx 1.2\times10^{14}$ gm/cm$^3$), where the neutron $^1S_0$ 
pairing presumably breaks up and where all nuclei have dissolved to nuclear 
matter, Eq.~(\ref{eq:tcmII}) can be used to set a lower-limit on 
the moment of inertia of this part of the star, $I_{icr}$. Taking the value 
obtained from the 1978 Vela glitch $\Delta\dot{\Omega}/\dot{\Omega}=0.024$
(\cite{Alparal}), a significant constraint is placed on the minimal size of 
the inner crust.

For any EOS, the larger the given gravitational mass, the more compact the 
neutron star and the larger the fraction of the mass held in the core. 
Both these effects combine to reduce the fraction of the moment of inertia 
of the inner crust, $I_{icr}/I_{tot}$, as the gravitational mass is increased. 
Hence, a larger observed value of $\Delta\dot{\Omega}/\dot{\Omega}$ 
implies a smaller value for the neutron star gravitational mass in order to 
satisfy the two-component condition of 
$I_{icr}/I_{tot}\geq\Delta\dot{\Omega}/\dot{\Omega}$. Roughly speaking, 
a similar combination occurs for a given gravitational mass when comparing 
a soft EOS to a stiff one, since the softer EOS will yield a more compact 
star and a smaller crust. We demonstrate this in Fig.~\ref{fig:Icr_vs_M}, 
which shows the fraction of the moment of inertia carried by the inner crust, 
$I_{icr}/I_{tot}$, as a function of mass for the nuclear matter equations 
of state presented in Fig.~\ref{fig:M_vs_R}, along with EOS 1 and 2 with all 
hyperon species. The moment of inertia of the neutron star was calculated in 
the slow-rotation approximation, again following Arnett and Bowers (1977). 

All equations show a decrease in  $I_{icr}/I_{tot}$ as a function of the 
gravitational mass, as discussed above. The key observation, however, is 
that, with the exception of the very stiff MF model, the inner crust 
of a $1.4\;M_\odot$ neutron star found for nuclear matter matter equations 
of state is too small to carry a moment of inertia with 
$I_{icr}\geq0.024I_{tot}$. The equations of state with hyperons are, on the 
other hand, able to support a large crust in a $1.4\;M_\odot$ star, in 
spite of their maximum masses being 
relatively low. This is because the crustal size mainly 
depends on the EOS of the matter just below it, i.e., at 
$\rho_B\!\approx\!\rho_0$, while the maximum mass is more sensitive to the 
EOS at higher densities. Thus, a large crust and a low maximum mass are 
easily reconciled for any equation of state that is stiff at lower 
densities, and softens at $\rho_B\!\geq\!2\rho_0$.

This may serve to indicate that the EOS should turn from stiff at low 
densities to soft at higher densities, as speculated in the previous
subsections. Again, hyperon formation offers a natural (but not unique) basis 
for such an EOS to prevail. Once more, we emphasize that this effect regarding 
crustal size is a generic feature of the inclusion of more 
species in dense supernuclear matter, and only the finer details 
will be model dependent.

%=================================================================
%    Cooling rates
%=================================================================

\section{Roles of Hyperons in Neutron Star Cooling Rates}

In recent years it has been possible to detect X-rays from over twenty 
pulsars (\cite{NStemp}). For a few pulsars, there is strong 
evidence that actual surface thermal radiation has been detected, while for 
others only upper limits can be stated. The surface temperature 
(interpreted from the surface radiation) and the pulsar age (usually 
estimated through the spin down rate) provide a constraint on the thermal 
history of the pulsar. Comparison of observation and theoretical models of 
neutron star cooling may offer a unique indication regarding the composition 
of the high density matter core, including the presence of new hadronic degrees
of freedom.

The implications of hyperon formation for neutron star cooling have been 
discussed in several recent studies (\cite{Prakalcool}, \cite{Prakcool}, 
\cite{Hanselcool}, \cite{Schaabcool}). The common theme of these works has been 
that hyperons provide additional channels for rapid cooling processes, i.e., 
the direct Urca. Hyperon direct Urca processes, like the nucleon direct Urca, 
are basically thermal fluctuations of baryons and leptons:
\begin{equation}
 B_1\rightarrow B_2+e+\bar{\nu}_e \;\; ; \;\; B_2+e\rightarrow B_1+\nu_e\;\; .
\end{equation}
The direct Urca processes allow for large neutrino emissivity, 
so that rapid core cooling dominates the star's thermal evolution. The direct 
Urca cooling emissivity, $\epsilon_{DU}$, has been estimated as 
(\cite{Prakalcool}) 
\begin{equation}\label{eq:eDU}
   \epsilon_{DU}=4\times10^{27}\left(\frac{x_e\rho_B}{\rho_0}\right)^{1/3}
            \frac{m_{B_1}m_{B_2}}{m_n^2}
                  R T_9^6\;\; \mbox{erg cm}^{-3} \mbox{s}^{-1} \;\;,
\end{equation}
where $x_e$ is the electron fraction per baryon, $m_{B_1}$ and $m_{B_2}$ 
are the effective masses of the two participating baryons, $m_n$ is the 
neutron mass, and $T_9$ is the core temperature in units of $10^9\;^oK$. 
$R$ is a weak interaction matrix element factor which is unity for nucleon 
Urca ($n\rightarrow p+e+\bar{\nu}_e$), and ranges between $\sim10^{-2}$ for 
strangeness changing reactions (such as $\Lambda\rightarrow p+e+\bar{\nu}_e$) 
and $\sim10^{-1}$ for strangeness conserving reactions which include 
hyperons (such as $\Sigma^-\rightarrow\Lambda\!+\!e\!+\!\bar{\nu}$);
see the review by Prakash (1994) for details.

If neutron stars cool through direct Urca processes indefinitely, 
their temperature should drop so rapidly that the surface temperatures 
(typically $10^{-2}$ of the core temperature) would be undetectable within 
less than 100 years of the star's birth (\cite{Latal94}). Observation seems 
to suggest otherwise, indicating that direct Urca processes are suppressed in 
the core through most of the star's thermal evolution, so that a 
significant surface temperature can be detected even at pulsar ages of 
$10^3-10^5$ years. If the direct Urca is indeed suppressed, then cooling
proceeds through less efficient processes, 
most of which have emissivities $\sim T_9^8$ 
(with various numerical coefficients (\cite{Max87})). Calculations of neutron 
star thermal evolutions in which slower cooling processes dominate do find that 
the surface temperature remains rather large for $\sim\!10^5$ years, 
until crust photon emission takes over as the dominant cooling process.
 
Direct Urca processes may be suppressed by two main mechanisms: absolute 
suppression if energy and momentum cannot be conserved, and partial
suppression if the participating baryons pair to a superfluid state.  
The presence of hyperons in neutron star cores has 
implications on both issues, as discussed below.

\subsection{\it Threshold Concentrations}   

Due to the extreme degeneracy of fermions in neutron star cores, direct Urca 
reactions take place only with baryons and leptons on their respective Fermi 
surfaces. Imposing energy and momentum conservation (and assuming a negligible 
neutrino energy: $E_{\nu}\!\approx\!k_BT$), leads to a combination of the 
conditions:
\begin{equation}\label{eq:Urcacon}
   \mu_{B_1}=\mu_{B_2}+\mu_e \;\;\;\; ; \;\;\;\;\ 
                              p_F(B_1)\leq p_F(B_2)+p_F(e) \;\;,
\end{equation}
$p_F(X)$ being the Fermi momenta of species X.
The first condition simply imposes chemical equilibrium which is fulfilled 
inherently (Eq.~(\ref{eq:genequi})), and the second is the 
``triangle inequality'', which must be fulfilled for all cyclic permutations 
of the $B_1,B_2$ and $e$. For Fermions $p_F\!=\!(3\pi^2x\rho_B)^{1/3}$, 
so that the second condition of Eq.~(\ref{eq:Urcacon}) becomes 
$x_1^{1/3}\!\leq\!x_2^{1/3}+x_e^{1/3}$ (again, along with all cyclic 
permutations). 

According to the momentum-conservation condition above, direct Urca processes 
may take place only if the relative fractions of the two baryon species are 
not too different from one another, and from the electron fraction as well. 
The fractions required set threshold conditions for the existence of direct 
Urca processes, which are otherwise completely extinct. 
For nuclear matter  in beta-equilibrium it has been shown that 
the large neutron excess requires a threshold proton fraction of at least 
$x_{pc}\!\geq\!11\!-\!15\!\%$ (\cite{Prakalcool}) to allow the nucleon direct
Urca process to take place. Whether or not a large enough proton fraction 
exists in the cores of neutron stars depends on the specifics of the nuclear 
matter EOS. Once hyperons appear in the matter, threshold concentrations are 
easier to meet for all types of direct Urca processes. First, the threshold 
concentrations for some hyperon processes with protons or other hyperons are 
inherently low - typically on the order of 0.01 - since their fractions are 
initially similar (\cite{Prakcool}).  
Second, since hyperon formation is followed by an increase of the proton 
fraction and a reduction of the neutron fraction, 
the threshold concentrations for neutron related Urca processes are 
also easier to fulfill, including nucleon direct Urca ($x_p/x_n\!\geq\!0.1$).

Direct Urca processes will become prohibited if the electron fraction is 
too small to allow for momentum conservation (i.e., the triangle inequalities 
cannot be fulfilled). However, for typical equilibrium compositions with 
hyperons, this happens only when the electron fraction drops below about 
$0.5\%$, which does not occur at the central density of a $1.4\;M_\odot$ star 
as found with practically all published equations of state.
The small lepton fraction induced by hyperon formation will also reduce the 
direct Urca emissivities through the $x_e$ dependence in Eq.~(\ref{eq:eDU}). 
However, even for $x_e\!\approx\!0.1\%$ this suppression is 
only by a factor of a few (and the effect on the cooling rate will be somewhat 
balanced by a reduction of the star's heat capacity). Hence, the composition 
of a hyperon rich core should allow for at least most direct Urca processes 
to dominate in typical neutron stars.  

\subsection{\it Superfluidity}
 
In view of the lenient conditions for hyperon direct Urca reactions, 
theoretical models of neutron star thermal evolution find that stars 
with hyperons will cool very rapidly (\cite{Hanselcool}, \cite{Schaabcool}). 
However, these analyses assumed that all hyperons are in a normal, rather
than a superfluid state. Here we wish to call attention to the implications of 
a recent estimate of hyperon pairing gaps.

Baryon superfluidity may have various consequences on neutron star properties
including a significant moderation of cooling processes. If the baryons on 
the Fermi surface couple to superfluid pairs with a gap energy of $\Delta$, 
the direct Urca emissivity is reduced by a factor of $\sim\exp(-\Delta/k_B T)$, 
since an energy of $\Delta$ is first required to break up the superfluid pair.
Nucleon pairing in neutron stars has received much attention, 
and for the last two decades it has been widely accepted that nucleons can 
couple to superfluid pairs in neutron stars. The commonly accepted picture 
(\cite{ShaTeu83}) is that neutrons in the inner crust couple to a $^1S_0$ 
superfluid, while in the core the neutrons couple to a $^3P_2$ superfluid 
(due to their high Fermi momenta) and the protons couple to a $^1S_0$ 
superconductor. Estimates of the gap energies have proven to be model 
dependent, but core gap energies are typically found to be in the range 
0.1 to 1 MeV. The existence of baryon pairing is
expected when the temperature drops below the critical temperature, which is 
$\sim\!0.57\Delta/k_B$ for $S-$wave pairing and $\sim\!0.12\Delta/k_B$ 
for $P-$wave pairing. Neutron stars are expected to cool below the critical 
temperatures for nucleon pairing within days after their birth, 
and so nucleon pairing is conventionally assumed to be present in 
neutron star cores, with a significant impact on nucleon direct Urca emissivity 
(\cite{Latal94}, \cite{Page95}, \cite{SchaabAA}).

Until recently, quantitative estimates of pairing of other baryon species 
have not been performed due to lack of relevant experimental data. 
In a recent work, Balberg \& Barnea (1998) used an analysis of doubly strange 
hypernuclei in a first attempt to determine pairing gaps for 
$\Lambda$ hyperons in a neutron star matter background. The $\Lambda$ hyperons 
were found to couple in a $^1S_0$ superfluid, with a gap energy 
of a few tenths of an MeV. $S$-wave pairing is expected for $\Lambda$ Fermi 
momenta up to about $1.3$ fm$^{-1}$, very much like the corresponding value 
for protons in a neutron matter background (\cite{Norvg}).
These results imply that for typical models of hyperon formation in neutron 
stars, a $\Lambda$ $^1S_0$ superfluid will exist between the threshold baryon 
density for $\Lambda$ formation and the baryon density where the $\Lambda$ 
fraction reaches $15\!-\!20\%$. While this result is based on limited data 
from double hypernuclei (i.e., nuclear matter background at normal nuclear 
density), the basic prediction of a $\Lambda$ superfluid is not surprising, 
in view of the general similarity of $\Lambda-\Lambda$ and 
nucleon$-$nucleon interactions. Further work is clearly necessary, and 
estimation of gap energies for other hyperon species is also required
(including possible anisotropic pairing modes), but in general, it seems 
prudent not only to allow for hyperon formation 
in neutron stars, but also to include hyperon superfluidity.

A full treatment of neutron star cooling with superfluid hyperons is beyond 
the scope of this study, and is reported to another work (\cite{SBScool}).
We point out that for suppression of all direct Urca processes, it is 
sufficient that only the neutral baryons, e.g., n, $\Lambda$ and $\Sigma^0$ 
(or n, $\Lambda$ and $\Xi^0$, if $\Sigma$'s are absent), couple to 
a superfluid state: charged baryons will be deprived of partners for the 
direct Urca processes. Furthermore, at central densities typical of a 
$1.4\;M_{\odot}$ neutron star (in most equations of state), fractions of 
neutral hyperons other than the $\Lambda$ are very small, so $\Lambda$ and 
neutron superfluidity are sufficient to significantly moderate all relevant 
direct Urca processes. The core temperature should then saturate according to 
the lower of the neutron $^3P_2$ and the $\Lambda$ $^1S_0$ critical 
temperatures, with the surface temperature declining very slowly 
for $10^4-10^5$ years. Correspondingly, hyperon formation can indeed be 
compatible with observed thermal emission from pulsars.

%=================================================================
%    Phase Transitions
%=================================================================
\section{Roles of Hyperons in Phase Transitions}

We now return to possible phase transitions of high density matter in the 
cores of neutron stars. There are two such transitions
which may occur in the matter: the formation of an $S$-wave
meson condensate, and deconfinement of baryons into quarks. Both types of 
transitions have been the subject of intensive study, but whether or not
one (or both) of them can actually take place in
cold, beta-equilibrated supernuclear density matter remains an open question. 
This is mainly because the details of the transitions are dependent on 
physical values (i.e., meson effective masses and quark matter physics) 
which are currently unattainable by experiment, leaving uncertainties that 
are too large to significantly constrain predictions. 

While these phase transitions are of obvious interest from the particle physics 
point of view, they also have astrophysical implications
through their possible effects on neutron star properties 
(mass radii relations, cooling rates, etc.). 
For our discussion here it is important that both types of transitions offer 
alternative hadronic degrees of freedom to hyperon formation. Indeed, both
meson condensation and deconfinement have been demonstrated to soften the
equation of state and cause deleptonization (offering negatively charged
hadrons or quarks to replace the leptons), therefore leading to most of the 
effects discussed in previous sections (\cite{PrakalPNS}). 
Furthermore, meson condensation and 
baryon deconfinement offer potential competition to hyperon formation, since 
they too lower the energy per baryon of the matter, thereby decreasing the 
nucleon chemical potentials. In this section we examine what influence the 
presence of hyperons in the cores of neutron stars can have on meson 
condensation and baryon deconfinement. 

\subsection{\it Meson Condensates}

Mesons may form freely in baryonic matter since they do not obey number 
conservation. At zero temperature, a nonzero meson density naturally takes
the form of a Bose condensate, where the required energy 
for meson accumulation is only the meson ground state energy which may be 
identified through the meson effective mass. Thus, the candidates considered 
most likely for condensation in neutron star matter are the lightest 
mesons - the pion and the kaon.

At low densities (including in nuclei), the available energy is insufficient to 
maintain a nonzero mesonic density
and mesons serve only as carriers of the baryonic 
interactions. At higher densities the available energy in the strong 
interactions increases, and at some finite density mesons may begin to 
condense. 

Since the baryonic content of the matter has a net positive charge,
the best candidates for condensation are the negatively charged mesons. 
The relevant meson creating reactions are: $B_1\rightarrow B_2+\pi^-$ or
$B_1\rightarrow B_2+K^-$ (with additional particles participating in order to 
conserve momentum). The fundamental point is that these processes 
are equivalent to the lepton-related reactions (Eq.~(\ref{eq:genpro})), so 
that the mesons basically compete with the charged leptons. Since neutrinos and 
anti-neutrinos are assumed to have zero chemical potential, the basic 
condition for negatively charged meson condensation is
\begin{equation}\label{eq:mM=mue}
                   m^*_{M^-}=\mu_e \;\;\; ,
\end{equation}
where $M^-$ denotes the $\pi^-$ or $K^-$, and $m^*$ denotes the meson effective 
mass, which may differ from the bare mass due to medium effects. Note that for 
neutral mesons the condition for condensation is $m^*_{M^0}\!=\!0$ and
for positively charged mesons it is even $m^*_{M^+}\!=\!-\mu_e$. 

\begin{figure}[htb]
\begin{center}
\figcaption{Fractional moment of inertia of the inner crust, $I_{icr}/I_{tot}$,
as a function of the static mass (in units of $M_\odot$) 
for EOS 1, EOS 2 (with and with out $\Sigma$ hyperons, marked as in
Fig.~\ref{fig:EOSS}), and the nuclear matter equations FPS, A, AU and L. The 
thin dashed horizontal line corresponds to the observational constraint of theVela 1978 glitch, $I_{icr}/I_{tot}\!\geq\!0.024$.
\label{fig:Icr_vs_M}}
\vspace{-2.0cm}
\includegraphics[width=10cm]{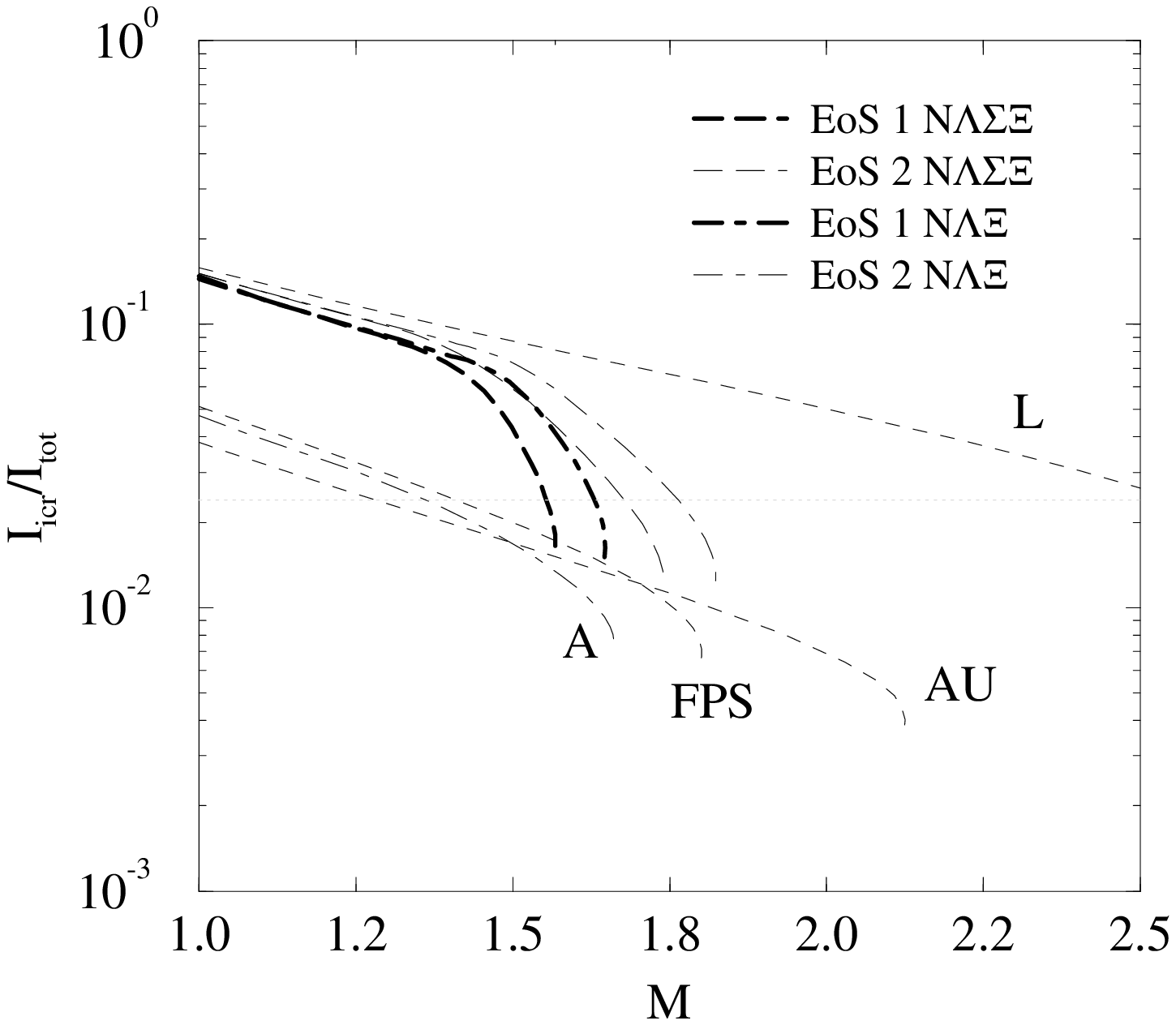}
\vspace{-2.0cm}
\figcaption{The electron chemical potential, $\mu_e$, for EOS 1 (thick lines) 
and EOS 2 (thin lines). The curves correspond to nuclear matter 
(solid line, identical for both equations), matter with nucleons and all 
hyperons (dashed lines) and matter with nucleons, $\Lambda$ and $\Xi$ 
hyperons but no $\Sigma$'s (dot-dashed lines).
\label{fig:mue_vs_rhoB}}
\vspace{-0.5cm}
\includegraphics[width=10cm]{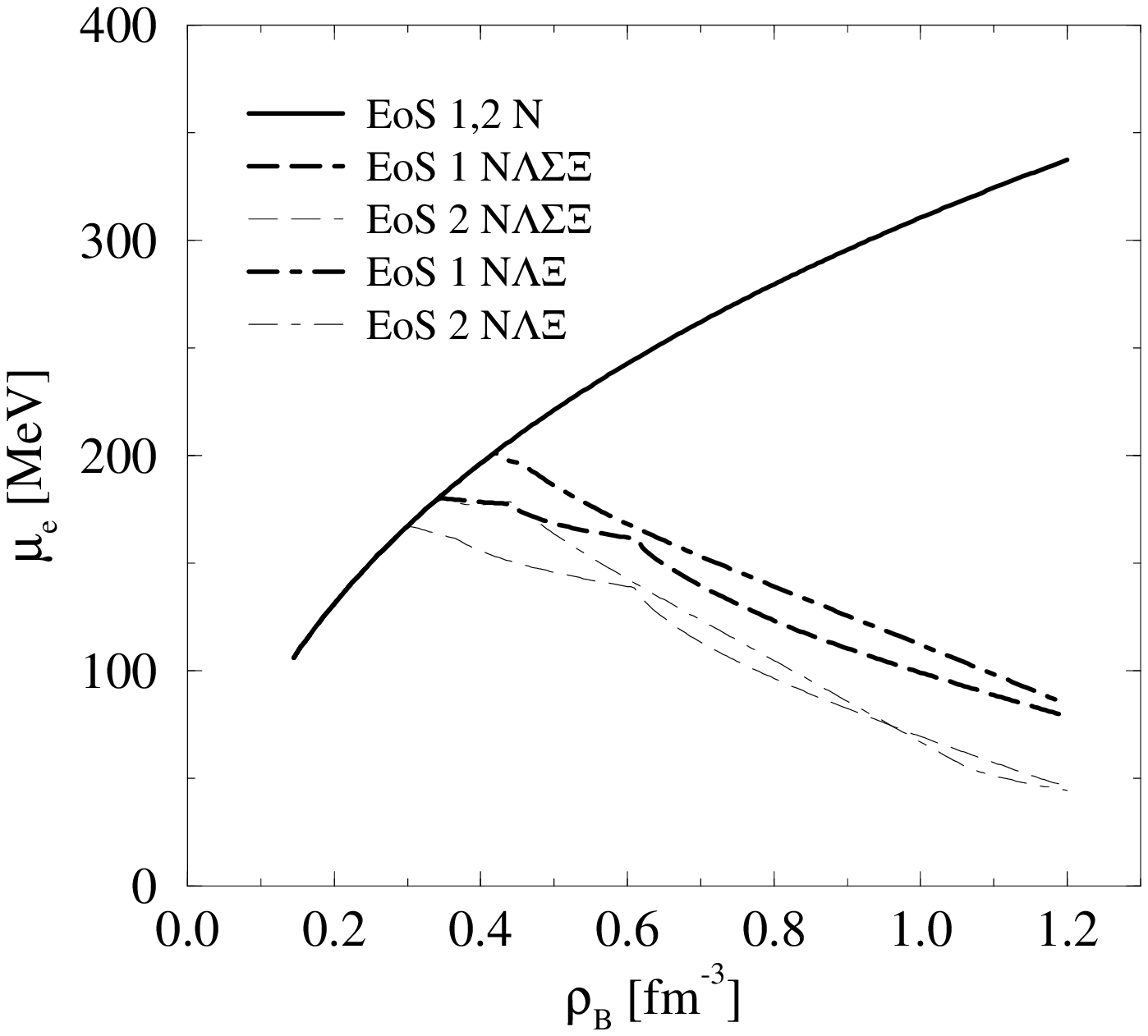}
\end{center}
\end{figure}

The condition in Eq. (\ref{eq:mM=mue}) implies that hyperon formation has 
a fundamental influence over the likelihood of meson condensation 
in neutron star matter. 
This can be seen explicitly in Fig.~\ref{fig:mue_vs_rhoB}, where the electron 
chemical potential is plotted as a function of baryon density for the 
equilibrium compositions found with equations of state 1 and 2 presented 
above. The plots correspond to the nuclear matter case (identical in both 
cases, since they have a common nuclear symmetry term), matter with 
all types of hyperons, and matter with $\Lambda$'s and $\Xi$'s but 
no $\Sigma$'s. 

The qualitative difference between nuclear matter and matter with hyperons 
is explicit, and is an obvious result of the deleptonization hyperons induce in 
the matter. For nuclear matter the electron fraction gradually rises for 
larger densities, and the electron chemical potential reaches 300 MeV and
more. On the other hand, the onset of hyperon accumulation is followed by a 
drop in the electron fraction, and a corresponding reduction in the electron 
chemical potentials. This is especially pronounced when
negatively charged hyperons appear. The electron chemical potential
typically reaches a maximum value which is somewhat model dependent at about 
200 MeV, and at very high densities drops to even less than 100 MeV.
Since limiting the electron chemical potential is an immediate consequence of
hyperon formation (see also \cite{Glenbook}), this suggests another general
result: meson condensation is less likely in matter with hyperons than in 
nuclear matter. 

Whether or not the deleptonization is sufficient to deny meson condensation 
also depends on the values of the meson effective masses, which are poorly 
known at present. Evaluating the meson effective mass as a function of the 
baryon density and composition requires the self-consistent inclusion of the 
meson fields in the Lagrangian, which we do not follow here. 

We do note that modern estimates of the $\pi$-nucleon interaction find that 
the $\pi$ effective mass is expected to grow with respect to the bare value of 
$\approx\!140\;$MeV due to the strength of the nucleon particle-hole 
interaction (\cite{Brn88}, \cite{Baym91}; see also Waas, 
Brockmann \& Weise 1997, for a recent estimate). Hence, even though the 
$\pi^-$ was considered a natural candidate for condensation in many early 
works, most current studies agree that such condensation is unlikely, even in 
nuclear matter. Hyperon formation will have an indirect effect on the 
likelihood of condensation of other pion species as well, since it reduces the 
total energy per baryon of the matter. However, this effect cannot be 
estimated quantitatively in the context of the models used in this work. In 
principle, repulsive $\pi N \Delta$ coupling should suppress condensation of 
other pion species as well (\cite{Baym91}).  But it should be noted that 
variational models performed with Argonne three-body forces 
(\cite{WFF}, \cite{APR98}) find that strong tensor correlations which imply 
neutral pion condensation appear in nuclear matter at very low 
densities of about $0.2\;$fm$^{-3}$.
 
Unlike the pion, analysis of $K^-$ atomic data (e.g., 
Friedman, Batty \& Gal 1994) suggests an attractive $K^-$-nucleon potential 
which reduces the bare mass of about $500$ MeV by a sizable 
fraction. Indeed, $K^-$ condensation in neutron star matter has become an 
active subject of investigation since first suggested by Kaplan and Nelson 
(1986). Some studies of $K^-$ condensation in {\it nuclear} matter have found 
a threshold condensation density as low as $3\!-\!4\rho_0$ (\cite{SBKaon}, 
Pandharipande, Pethick \& Thorsson, 1995). However, an effective $K^-$ mass as 
low as 200 MeV at a density of $2\rho_0$ seems unattainable in present studies 
(see also Schaffner et al. 1994). In view of the analysis displayed in 
Fig.~\ref{fig:mue_vs_rhoB}, this implies that hyperons will appear in nuclear 
matter prior to the onset of $K^-$ condensation, and thus delay the 
condensation to higher densities. This observation is 
indeed supported by the results of two recent studies 
(\cite{SBH}, \cite{SMH}), which examined $K^-$ condensation along with hyperon 
formation. Both find that hyperons appear before the $K^-$ can condense in 
nuclear matter, and that condensation in matter with hyperons is then delayed 
to very large densities ($\geq\!8\rho_0$) or even completely suppressed. 

Clearly, should meson condensation occur at lower densities than hyperon 
formation, it is the latter that will be delayed, but this alternative seems 
less likely in view of most current works. Assuming it is hyperon formation 
that precedes, it will increase the threshold density for meson condensation, 
and this is once again a general feature of allowing for hyperon formation in 
neutron stars.

\subsection{\it Baryon Deconfinement}

The possibility of a phase transition of high density baryonic matter into
quark matter through baryon deconfinement has received much attention both 
in the context of neutron stars and in heavy ion physics. Intuitively, it 
seems inevitable that at large enough densities the quarks will no longer
retain their arrangement as baryons, but rather deconfine into larger ``bags''
of quarks, and eventually into quark matter. However, present theoretical 
limits and uncertainties in the modeling of QCD prevent a comprehensive 
analysis of deconfinement physics. Current studies rely on simplified 
models, and published results and conclusions prove to be highly model 
dependent. Nonetheless, as we show below, the presence of hyperons does have 
a general effect on the phase transition between baryonic and quark matter.

Baryon deconfinement in high density matter is expected to 
proceed gradually with increasing density through a continuous mixed phase
with various spatial combinations of the two phases. This is analogous to the 
transition from nuclei to nuclear matter in the inner crusts of 
neutron stars. The ground state at each density is achieved by arranging the 
composition, density and shape of each phase, including  long range 
ordering enforced by the Coulomb interaction. This character of the transition
from baryon matter to quark matter was pointed out by Glendenning (1992, 
see \cite{Glenbook} for a review), and is a natural consequence of the presence
of two conserved charges - baryon number and electric charge. Pressure 
varies continuously over the density range of the transition, rather
than remaining constant as in a ``standard phase transition''. A neutron star 
which includes a mixed phase (and possibly, at very high densities, a pure 
quark phase) is often referred to as a ``hybrid star''. 

It is important to note that the conditions for the onset of deconfinement 
differ considerably from the conditions for equilibrium of the two phases. 
The key factor is the difference in the strangeness fraction of the equilibrium 
compositions of the two phases. In the quark phase the difference between the 
mass of the strange quark and those of the up and down quarks is significantly 
lower than the Fermi energies of the quarks. The equilibrium quark 
composition should therefore hold almost equal fractions of the three flavors, 
with a strangeness fraction per baryon of almost unity. 
In the baryonic phase an overall strangeness fraction of unity can be reached 
only at very high densities, when the baryon masses no longer dominate the 
values of the chemical potentials. 

Once a stable quark matter phase is created, it reaches its equilibrium 
composition through weak decays, regardless of the initial baryon 
composition.  The initial deconfinement process, however, must 
take place through the strong interaction, which conserves flavor. The minimum 
energy state a quark phase component can achieve through deconfinement 
will not be its ground state, but only the lowest energy state with the 
available underlying quark composition of the baryons. It is therefore 
straightforward that when the baryonic matter includes a finite strangeness 
fraction (i.e., hyperons are present), the energy a quark phase component 
can achieve through deconfinement will be lower than for a two-flavor quark 
phase created by deconfinement of pure nucleon matter. 
The presence of hyperons lowers the energy per baryon with respect to nuclear 
matter in the baryonic phase as well, but the effect of a finite strangeness 
fraction in the quark phase is significantly larger. The threshold for 
deconfinement of matter with hyperons should thus be lower than for nuclear 
matter.

\begin{figure}[htb]
\begin{center}
\figcaption{Energy per baryon, $E/A$, for baryonic matter (thick lines) and for 
quark matter of identical composition (thin lines) as a function of the 
baryonic density. The baryonic matter is calculated with EOS 1 
(and the composition corresponds to the equilibrium composition of this 
equation). The quark matter is calculated with the MIT bag model with 
(a) $B=100\;$ Mev fm$^{-3}$, $\alpha_c=0$ 
(b) $B=70\;$ Mev fm$^{-3}$, $\alpha_c=0.3$. The solid lines are nuclear 
matter and the corresponding two-flavor quark matter; the dashed lines are the
baryonic matter with hyperons and the corresponding three-flavor quark matter.
\label{fig:E/A_vs_rhoB}}
\includegraphics[width=10cm]{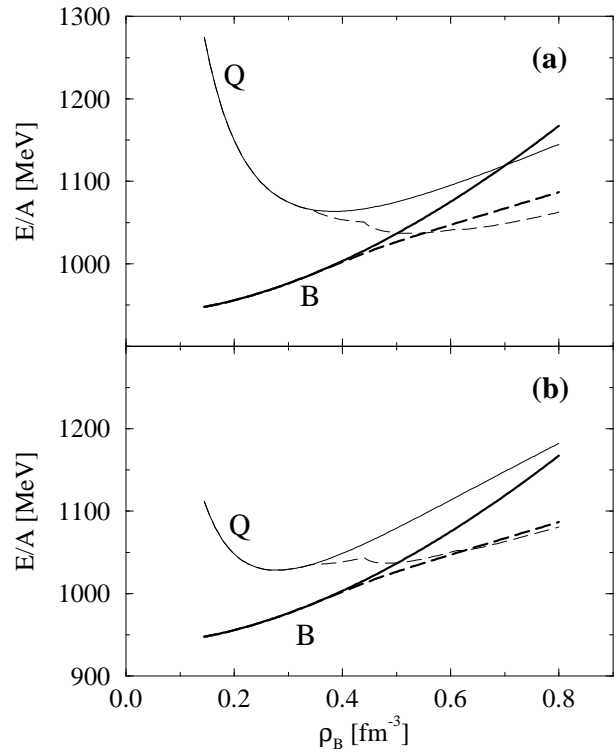}
\end{center}
\end{figure}

We demonstrate this qualitative description with a crude analysis for the 
deconfinement threshold, as follows: assume that baryonic matter 
deconfines as a bulk, where baryons at a given density $\rho_B$ 
deconfine spontaneously to quark matter of identical density and quark 
composition. Such deconfinement 
will proceed if the energy per baryon in the quark phase is equal to (or less than)
that in the baryonic phase. Figure~\ref{fig:E/A_vs_rhoB} compares the energy 
per baryon, $E/A\!=\!\varepsilon_B/\rho_B$, as a function of the density of the 
equilibrium baryon compositions of EOS 1 with the energy per baryon of quark 
matter of identical composition and density. The properties of the quark phases 
in each case are calculated with simple MIT bag model parameterizations, 
where $B$ is the bag constant (in MeV/fm$^3$) and $\alpha_c$ is the strong 
interaction coupling constant (see chapter 8 in \cite{Glenbook}).

The baryon density of deconfinement corresponds to the 
crossover between the baryonic matter and quark matter curves. The effect of 
finite strangeness on the energy per baryon is much more 
pronounced in the quark phase, resulting in an observable decrease of 
the energy per baryon when a new hyperon species appears 
in the matter (i.e., where the strangeness fraction increases rapidly with 
density). The density of deconfinement for any given quark matter model is 
then lower for matter with hyperons than for nuclear matter. 

The specific value of the deconfinement density is strongly dependent on 
the values of the bag model constants. Since the values  used in the 
calculations are arbitrary, emphasis is placed on the fundamental reduction 
in the threshold for deconfinement in the presence of hyperons. 
Once again, this
is a general feature of allowing hyperons to appear in the matter. We also find 
a lower deconfinement density for other high density equations of state 
(when comparing matter with hyperons and nuclear matter), and that it is
independent of the specific choice of quark bag model constants.

While the assumption that the baryon-quark phase transition proceeds through 
bulk deconfinement is a crude one, it actually provides 
an upper limit for the deconfinement threshold. The alternative 
to bulk deconfinement is nucleation of quark bubbles in the baryonic 
background, when only some of the baryons deconfine. For a quark bubble to 
survive it must maintain thermodynamic and chemical 
equilibrium of strong interactions with the baryon background while the net 
flavor of all of the matter must still be conserved. Unlike the 
bulk deconfinement scenario, quark bubbles can compress or expand to 
have a different density than the surrounding medium. They can also maintain 
a composition different from the original baryons by controlling the 
fractions of baryon species which deconfine. This implies that bubble 
nucleation has more degrees of freedom than bulk deconfinement, and the 
possibility for the quark bubbles to have a larger strangeness 
fraction than the baryon background is especially helpful for bubble formation. 
This qualitative argument suggest both that the threshold density for 
deconfinement is lower for bubble nucleation than for bulk deconfinement, 
and that the impact of strangeness being present should be even more 
pronounced in this case. Calculations of bubble nucleation are, however, 
highly model dependent, since additional factors such as the bag surface 
tension must be included (see, e.g., \cite{OlMadDC}), and we confine ourselves 
to these qualitative remarks. 

Finally, it is interesting to note that in equilibrium, the presence 
of hyperons is expected to increase the minimal baryon density in which 
a mixed phase can exist with respect to nuclear matter (\cite{PrakalPNS}). 
Hyperon formation reduces the energy per baryon, and at any total 
baryon density the quark component of the mixed phase always 
occupies less volume and holds less baryon number when equilibrated with 
a baryonic phase with hyperons than when equilibrated with nucleons. However, 
since the mixed phase can only appear following the initial 
deconfinement, hyperon formation {\it does not} suppress deconfinement, but 
rather enhances it, as is evident from Fig.~\ref{fig:E/A_vs_rhoB} and the 
discussion above. 
Furthermore, since hyperons soften the equation 
of state, a larger central baryon density is required to support a star of given
mass than for nuclear matter. We conclude that the likelihood of deconfinement 
and creation of a mixed phase in neutron stars is {\it increased} by the 
appearance of hyperons with respect to nuclear matter. This likelihood is 
difficult to quantify, since it is strongly dependent on the quark matter 
equation of state, which may delay deconfinement to extremely high densities,
or even forbid it from occurring in the density range relevant to 
neutron star cores (in the context of the bag model, this is naturally 
achieved by increasing $B$). 
Nonetheless, these general implications of hyperon formation are, 
yet again, model independent.

%\pagebreak
%=================================================================
%   Conclusions and Discussion
%=================================================================
\section{Conclusions and Discussion}

Various recent studies of hyperon formation in neutron stars 
share a consensus, that hyperons will appear in the cores of neutron stars
at a density of about $2\rho_0$. This consensus is a direct consequence of
the common basis used in these works for describing the hyperon-nucleon 
and hyperon-hyperon interactions, as deduced from hypernuclei experiments.
The qualitative principle which arises from these experiments is that 
hyperon related interactions are similar both in character and in order 
of magnitude to the nucleon-related interactions. It is thus reasonable for the 
fundamental similarity of nucleons and hyperons to be manifest at higher 
densities, where the typical energy scales are of the order of the mass
differences between the different species.
In this study we concentrate on 
the effects hyperon formation may have on the global properties of neutron 
stars. We discuss the implications of the presence of hyperons in the cores 
of neutron stars for the high density equation of state, for cooling 
reactions, and for phase transitions which could be possible in high density 
matter.

The fundamental effect of hyperon formation on high density matter is the 
softening of the equation of state with respect to the equation found 
for nuclear matter when using otherwise identical assumptions regarding 
the strong interactions. This effect is found in all the works 
which include hyperons in high density matter, and simply reflects that a 
larger number of baryonic degrees of freedom relieves some of the Fermi 
pressure of the nucleons. We also demonstrate a more subtle effect, where the 
rate at which the strong interaction potential energy density rises as a 
function baryonic density (due to repulsive short range forces) determines both 
the stiffness of the nuclear matter equation of state and the rate of hyperon 
accumulation. These two processes tend to balance one another in terms of the 
overall equation of state, and so hyperons 
induce a ``pressure control'' mechanism in the matter, in the sense that 
the equations of state for matter with hyperons are limited to a narrower range 
than nuclear matter equations of state. The clearest manifestation of the 
``pressure control'' mechanism is that the maximum neutron star masses found 
for equations of state with hyperons are limited to a rather narrow range - 
in this work $1.5\!-\!1.8\;M_\odot$ - much smaller than the range found 
for nuclear matter equations. A narrow range for the value of the maximum 
mass is in good agreement with published works, and we emphasize that it is 
a fundamental consequence of hyperon formation in neutron stars, while 
specific details of the modeled interactions are only of secondary importance.

Hyperon formation provides a natural route to combine a stiff equation of 
state at a density of $\rho_B\!\sim\!\rho_0$ and a softer equation at higher 
densities $\rho_B\!\geq\!2\rho_0$. Such a combination could make a large 
mass neutron star significantly more compact than a $1.4\;M_\odot$ star, and 
can even enable some specific configurations to 
undergo a spin-up period during their rotational evolution. 
Current observations do not provide positive or negative indication regarding 
such a combination in the equation of state, but we do point out that it can 
also reconcile a large neutron star crust, implied from 
pulsar glitch phenomena, with a low to intermediate maximum mass.
This is in contrast to the argument made regarding nuclear equations of state,
claiming the glitches indicate a stiff equation at high densities.

Once hyperons are included, the baryon composition of neutron star cores 
provides the necessary concentration thresholds for direct Urca neutrino 
emitting processes. This can lead to rapid cooling rates of neutron stars, 
but the actual cooling rates may be severely moderated if the baryons couple 
to superfluid pairs. While nucleon superfluidity has been discussed 
extensively in the past, we point out that hyperons should also be expected 
to be in a superfluid state, with gap energies in the same order of magnitude. 
Hence, hyperon formation can be consistent with observed cooling rates.

Both meson condensation and baryon deconfinement offer alternative degrees 
of freedom to hyperon formation in high density matter. The threshold 
densities for both types of phase transitions are difficult to constrain, 
due to large uncertainties involved. However, current estimates of the 
properties of high density matter suggest that hyperon formation will precede 
meson condensation and deconfinement. We show that if indeed hyperons appear 
first in the matter they affect the likelihood of both these transitions. 
Meson condensation is suppressed, since hyperons induce deleptonization, 
thus lowering the lepton chemical potential with which the mesons compete. 
On the other hand, the finite strangeness fraction allows the baryons to 
deconfine into a lower energy quark matter state, making deconfinement more 
favorable for any given quark matter physics. Both these trends are opposite 
than those expected in high density nuclear matter, which has a large 
lepton fraction and is composed of only strangeless quarks.

We emphasize again that all these results are basically general features of 
matter with multiple baryon species. Most of the specific quantitative values 
are dependent on the details of the modeling of the strong interaction, 
but the qualitative results discussed here should prevail as long as 
the general nature of these models is similar. It should be borne in mind that 
while the extrapolation of these interactions from hypernuclei data seems 
reasonable, large uncertainties still remain. If some of the hyperon-nucleon 
interactions are critically different than those assumed here, the qualitative 
trends can change through suppressing the formation of some baryon species. 

Unfortunately, the current status of observations of neutron stars does
not provide any significant constraints on the properties of high density matter.
Specific indications for the presence or absence of hyperons are
naturally unavailable as well. Clearly, any further measurements of 
masses, radii and rotation frequencies will be extremely valuable for 
constraining the high density equation of state, especially if unusual values 
(such as a large mass or a rapidly rotating star) will be observed. 
Correlations between these different quantities for any given object are also 
important, and the newly discovered Quasi-Periodic-Oscillations in kilohertz 
emission in some X-ray binaries may offer potential in this regard. 

Unique features of neutron stars, if observed, may also provide some 
indication regarding the physics of high density matter. If several hyperon 
species do form in the core, the core could be composed of multiple 
superfluids, including some negatively charged superconductors. Such a 
composition might have effects on both the rotational properties and the 
magnetic evolution of the star. Further differences with respect to nuclear 
matter might be found due to the lower density (by more than an order of 
magnitude) of the normal (not-superconducting) lepton component, in 
particular through the magnetic field evolution.  

Needless to say, additional experimental data from hypernuclei will be useful 
in establishing the foundations of high density matter models. This is 
especially relevant to the hyperon-nucleon interactions, for which relevant 
systems are more likely to be produced in current accelerators than  
for hyperon-hyperon interactions.

Finally, we recall that the properties of high density matter may have 
important consequences in several related astrophysical processes. Of these, 
the evolution of a newly born neutron star has received extensive attention 
in recent years (see, e.g, \cite{KJ}, \cite{PrakalPNS}). A unique qualitative 
feature in this regard is that matter with hyperons 
(and also with other negatively charged hadrons or quarks) will support a 
smaller maximum mass after neutrinos diffuse from the newly born core than 
while neutrinos are still trapped. Once lepton number in the matter is allowed 
to decrease, more hadronic degrees of freedom can be exploited which will
soften the equation of state. For nuclear matter the opposite occurs, since 
deleptonization leads to a larger neutron-proton asymmetry, which stiffens 
the equation of state. The maximum mass of a star with hyperons in the core is 
found to be larger when neutrinos are still trapped than after 
deleptonization, implying a mass range for which a newly born neutron star is 
meta-stable. Hyperon formation may thus play a role in creating another route 
for the formation of low mass black holes in type II supernova, 
with neutrino emission setting the time scale for the collapse (10-15 seconds). 
Such a scenario is especially appealing in view of the neutrino measurements 
from SN1987A (Ellis, Lattimer \& Prakash 1996), 
in which no neutron star has been found. If a 
future nearby supernova will provide finer details regarding the
emitted lepton number, neutrino energy content and time structure, 
valuable information may be 
inferred regarding high density matter and its composition, and hyperon 
formation in particular.

\acknowledgments
We are grateful to Avraham Gal, Zalman Barkat and Christoph Schaab for valuable 
advice and suggestions. This research was partially supported by the 
U.S.-Israel Binational Science Foundation grant 94-68.
\vspace {0.2in}

\newpage
\appendix
\section{The Effective Equation of State}

The formalism of the effective equation of state discussed in this work was 
presented by Balberg \& Gal (1997, to which the reader is referred for 
detail). This equation is basically a generalization of the Lattimer-Swesty 
equation of state for nuclear matter (\cite{LS}), which is commonly used in 
hydrodynamical simulations. An effective equation of state does not presume 
to describe the underlying physics of the strong interactions, but does allow 
for conducting extensive parameter surveys (including finite temperatures). 
The effective equation was shown to reproduce the main results of field 
theoretical models in terms of the equilibrium compositions and 
the thermodynamic properties of the high density matter.

The core of the effective EOS is an adjustable baryon-baryon 
potential, which models the strong interaction and provides the quantitative
description of the potential energy density of high density matter. This 
energy density is added to the kinetic energy and mass densities of the 
baryons, and to the kinetic energy and mass densities of the leptons, which 
are taken to be noninteracting (as is commonly assumed). 

The baryon-baryon interactions among the various species are described by 
assuming local density-dependent potentials. These potentials are constructed 
to reproduce the basic features of the strong interactions, i.e., long-range 
attraction and short range repulsion, and - in some cases - charge dependence, 
compatible with isospin invariance. The potential felt by a single baryon of 
species $y$ in bulk matter of baryon species $x$ with number density $\rho_{x}$ 
is then given in the form
\begin{equation} \label{eq:genlocpot}
   V_y (\rho_{x}) = a_{xy}\rho_x + 
                    b_{xy}t_x t_y \rho_{x} + 
                    c_{xy}\rho_x^{\gamma_{xy}}+
                    w \rho_x^\theta  \;\;.
\end{equation}
The first term yields attraction ($a_{xy}$ negative), the third and fourth 
terms yield repulsion ($c_{xy}$ and $w$ positive, supposedly introducing 
multibody interactions), and the second (symmetry) term introduces the charge 
dependence through a charge (isospin) $t$. Both $\gamma_{xy}$ and $\theta$ 
are greater than unity so that repulsion will dominate at high densities 
(short ranges). The fourth term included here was found to preempt 
some numerical problems which arose when a single repulsive term was used 
(\cite{BGH}). This term is assumed to represent universal short range 
interactions, thus setting $w$ and $\theta$ to be independent of the 
baryon species involved. The values of the coefficients and exponents are 
chosen to reproduce experimental data and accepted theoretical results 
regarding the baryon-baryon interactions. 

The local potential for a single baryon in a bulk of other baryons may be 
extrapolated into the potential energy density of bulk matter with a total 
density $\rho$ that includes both types of baryons. This is done by folding 
$V_y(\rho_x)$ with the partial density $\rho_y$, and vice versa 
($V_x(\rho_y)$ with $\rho_x$), combined with weight factors 
(avoiding double counting of each interaction). For simplicity we assume 
that all baryon-baryon interactions have a common value of $\gamma$ (further 
generalization is, of course, possible), and follow the common assumption 
of universal hyperon couplings (denoted below as $a_{YY}$ and $c_{YY}$). The 
final expression for the potential energy density of baryonic matter with a 
baryon density $\rho$ is then:
%\pagebreak

\begin{eqnarray} \label{eq:epotbar}
   \varepsilon_{pot}(\rho) = \!\!& &\!\!
   \frac{1}{2}\left[a_{NN}N^2\rho^2+b_{NN}(n-z)^2\rho^2+
                     c_{NN}N^{\gamma+1}\rho^{\gamma+1}\right]  \\ 
   \!\! & & \!\!\!\!+\:a_{\Lambda N}N\Lambda\rho^2+
       c_{\Lambda N}\left(\frac{N}{N+\Lambda}N^\gamma\Lambda+
       \frac{\Lambda}{N+\Lambda}\Lambda^\gamma N\right)\rho^{\gamma+1} 
                                                                 \nonumber \\
   \!\! & & \!\!\!\!+\:\frac{1}{2}\left[a_{\Lambda\Lambda}\Lambda^2\rho^2
                    +c_{YY}\Lambda^{\gamma+1}\rho^{\gamma+1}
   +a_{YY}\Xi^2\rho^2+b_{\Xi\Xi}(\Xi^--\Xi^0)^2\rho^2+
            c_{YY}\Xi^{\gamma+1}\rho^{\gamma+1}\right]     \nonumber \\
   \!\! & & \!\!\!\!+\:a_{\Xi N}N\Xi\rho^2+b_{\Xi N}(n\!-\!z)(\Xi^-\!-\!\Xi^0)
   \rho^2+c_{\Xi N}\left(\frac{N}{N+\Xi}N^\gamma\Xi+
       \frac{\Xi}{N+\Xi}\Xi^\gamma N\right)\!\rho^{\gamma+1}
                                                                 \nonumber \\
   \!\! & & \!\!\!\!+\:a_{YY}\Xi\Lambda\rho^2+
       c_{YY}\left(\frac{\Lambda}{\Xi+\Lambda}\Lambda^\gamma\Xi+
       \frac{\Xi}{\Xi+\Lambda}\Xi^\gamma \Lambda\right)\rho^{\gamma+1} 
                                                                 \nonumber \\
   \!\! & & \!\!\!\!+\:a_{\Sigma N}N\Sigma\rho^2+b_{\Sigma N}(n\!-\!z)
   (\Sigma^-\!-\!\Sigma^+)\rho^2+c_{\Sigma N}\!\left(\frac{N}{N+\Sigma}N^
   \gamma\Sigma+\frac{\Sigma}{N+\Sigma}\Sigma^\gamma N\right)\!\rho^{\gamma+1} 
                                                                 \nonumber \\
   \!\! & & \!\!\!\!+\:a_{YY}\Sigma\Lambda\rho^2+
       c_{YY}\left(\frac{\Sigma}{\Sigma+\Lambda}\Sigma^\gamma\Lambda+
       \frac{\Lambda}{\Sigma+\Lambda}\Lambda^\gamma\Sigma\right)\rho^{\gamma+1} 
                                                                 \nonumber \\
   \!\! & & \!\!\!\!+\:a_{YY}\Sigma\Xi\rho^2+
              b_{\Sigma \Xi}(\Xi^-\!-\!\Xi^0)(\Sigma^-\!-\!\Sigma^+)\rho^2+
       c_{YY}\!\left(\frac{\Xi}{\Xi+\Sigma}\Xi^\gamma\Sigma+
       \frac{\Sigma}{\Xi+\Sigma}\Sigma^\gamma \Xi\right)\!\rho^{\gamma+1}
                                                                 \nonumber \\
   \!\!& & \!\!\!\!+\frac{1}{2}\left[a_{YY}\Sigma^2\rho^2+
                        b_{\Sigma\Sigma}(\Sigma^--\Sigma^+)^2\rho^2+
       c_{YY}\Sigma^{\gamma+1}\rho^{\gamma+1}\right]   
   +\frac{1}{2}w\rho^{\theta+1}   \nonumber  \;,
\end{eqnarray}
where for each species $x=\rho_x/\rho$ and introducing the shortened notation
$N\!=\!n+z$; $\Xi\!=\!\Xi^0+\Xi^-$; $\Sigma\!=\!\Sigma^++\Sigma^0+\Sigma^-$.
In equation (\ref{eq:epotbar}) it is assumed that isospin (symmetry) forces 
vanish for the $\Lambda$ and $\Sigma^0$, which have a zero isospin projection
($I_3\!=\!0$).

Current experimental data are insufficient to constrain all the presented 
coefficients, especially those corresponding to the short range interactions.  
Once the shape of these interactions is assumed (by setting the value of $w$ 
and $\theta$), the rest of the coefficients can be determined by fitting them 
to reproduce the properties of nuclei and hypernuclei. In this work two 
sets of coefficients were used, denoted models 1 and 2. These models differ 
mainly in their density dependence (the values of $\gamma$ and $\theta$), 
and correspondingly reflect two values for the incompressibility of symmetric
nuclear matter at $\rho_N\!=\!\rho_0$: (1) $K\!=\!240\;$MeV, which is a 
``standard'' value used for high density matter, and (2) $K\!=\!320\;$MeV 
which implies a stiff EOS. The values of the coefficients in both 
equations are given in Tab. \ref{tab:eoscoef}. In the variation where $\Sigma$ 
hyperons are excluded from the matter, the coefficients of the $\Sigma$ related 
interactions are ignored.

Since the effective EOS has no means of consistently combining relativistic
and medium effects, the masses are set to be equal to the bare ones. This is 
a somewhat crude approximation, since field theoretical models suggest an 
effective baryon mass lower than the bare one, although the values of the 
coefficients in the potential energy expression can compensate in part for 
this approximation. Correspondingly, the baryons are treated as 
non-relativistic. We note that the equations do reach the causality limit, 
$dP/d\varepsilon\!=\!c^2$, but in both cases this occurs at densities which 
correspond to neutron star masses slightly higher than the maximum static mass. 

Approximating neutron star matter to have zero temperature, the kinetic 
energy and mass density terms for the baryons are, respectively:
\begin{equation}
\varepsilon_{kin}(\{x_i\},\rho_B)=\sum_{i}\frac{p_F^2(x_i)}{2m_i}\;\;\;\;
\varepsilon_{mass}(\{x_i\},\rho_B)=\sum_{i}x_i\rho_B m_i\;\;\; ,
\end{equation}
where $p_F(x_i)\!=\!\hbar c(3\pi^2x_i\rho_B)^{1/3}$ is the Fermi momenta of the 
baryons. The lepton relativistic 
energy densities (dependent on the lepton fraction and the total density) 
are added as well, yielding the final EOS, $\varepsilon(\{x_i,x_l\},\rho_B)$.

\newpage

%\begin{center}
%{\large Figure Captions}
%\end{center}

\newpage
\vspace{-10.0cm}
\newpage
\vspace{-10.0cm}

\newpage
\vspace{-10.0cm}

\newpage
\vspace{-10.0cm}

\newpage

\pagebreak

\newpage
\setlength{\parskip}{0.1in}
\setlength{\parindent}{0.0in}
%========================================================================
\begin{deluxetable}{l r r r}
\tablecaption{Quantum numbers of the baryons in the spin $\frac{1}{2}$ octet: 
mass (in MeV/$c^2$), strangeness, and isospin projection \label{tab:baryons}} 
\tablewidth{0pt}
\tablehead{ 
  & \colhead{mass} & \colhead{$S$} & \colhead{$I_3$} 	 }
\startdata
p          &  938.3      &  0   & $\frac{1}{2}$  \nl
n          &  939.6      &  0   & $-\frac{1}{2} $ \nl
$\Lambda$  & 1115.6      &  -1  & $  0         $ \nl
$\Sigma^+$ & 1189.4      &  -1  & $  1         $ \nl
$\Sigma^0$ & 1192.5      &  -1  & $  0         $ \nl
$\Sigma^-$ & 1197.3      &  -1  & $  -1        $ \nl
$\Xi^0$    & 1314.9      &  -2  & $\frac{1}{2}$ \nl
$\Xi^-$    & 1321.3      &  -2  & $-\frac{1}{2} $ \nl
\enddata
\end{deluxetable}

%========================================================================
%\clearpage
\begin{deluxetable}{lcclcc} 
\tablecaption{Coefficients for the potential energy density term in Eq. 
(\ref{eq:epotbar}) \label{tab:eoscoef}}                 
\tablewidth{0pt}
\tablehead{
   & \colhead{EOS 1} & \colhead{EOS 2} &  & \colhead{EOS 1}  & \colhead{EOS 2}}
\startdata
  $K$ [MeV]                                  &   240   &  320 
& $a_{\Sigma N}$\tablenotemark{\dagger}      & -481.3  & -354.8  \nl
  $\gamma$                                   &   4/3   &  5/3    
& $b_{\Sigma N}$\tablenotemark{\dagger}      &  214.2  &  214.2  \nl
  $\theta$                                   &   5/3   &   2
& $c_{\Sigma N}$\tablenotemark{\ddagger}     &  499.6  &  484.3  \nl
  $w$\tablenotemark{\star}                   &  223.6  &   220
& $a_{\Xi N}$\tablenotemark{\dagger}         & -410.2  & -303.1  \nl
  $a_{NN}$\tablenotemark{\dagger}            & -690.0  & -481.7
& $b_{\Xi N}$\tablenotemark{\dagger}         &    0    &    0    \nl
  $b_{NN}$\tablenotemark{\dagger}            &  107.1  &  107.1
& $c_{\Xi N}$\tablenotemark{\ddagger}        &  415.3  &  394.6  \nl
  $c_{NN}$\tablenotemark{\ddagger}           &  744.6  &  715.5      
& $a_{YY}$\tablenotemark{\dagger}            & -676.1  & -513.3  \nl         
  $a_{\Lambda N}$\tablenotemark{\dagger}     & -481.3  & -354.8
& $b_{\Sigma\Sigma}$\tablenotemark{\dagger}  &  214.2  &  214.2  \nl
  $b_{\Lambda N}$\tablenotemark{\dagger}     &    0    &    0
& $b_{\Xi\Xi}$\tablenotemark{\dagger}        &    0    &    0    \nl
  $c_{\Lambda N}$\tablenotemark{\ddagger}    &  499.6  &  484.3
& $c_{YY}$\tablenotemark{\ddagger}           &  658.1  &  764.7  \nl \tableline
\tablenotetext{\dagger}{MeV fm$^3$}
\tablenotetext{\ddagger}{MeV fm$^{3\gamma}$}
\tablenotetext{\star}{MeV fm$^{3\theta}$}
\enddata
\end{deluxetable}
%========================================================================
\end{document}